# A Practical Evaluation of Commercial Industrial Augmented Reality Systems in an Industry 4.0 Shipyard

**ÓSCAR BLANCO-NOVOA[1], TIAGO M. FERNÁNDEZ-CARAMÉS[1], (Senior Member, IEEE), PAULA FRAGA-LAMAS[1], (Member, IEEE), and MIGUEL VILAR-MONTESINOS[2]**

[1] Unidad Mixta de Investigación Navantia-UDC, Universidade da Coruña, Edificio de Talleres Tecnológicos, Mendizábal, s/n, 15403, Ferrol, Spain (e-mail: o.blanco@udc.es; tiago.fernandez@udc.es; paula.fraga@udc.es)

[2] Navantia S. A., Unidad de Producción de Ferrol, Taxonera, s/n, 15403, Ferrol, Spain. (e-mail:mvilar@navantia.es)

Corresponding authors: Tiago M. Fernández-Caramés and Paula Fraga-Lamas (e-mail: tiago.fernandez@udc.es; paula.fraga@udc.es).

This work is part of the Plant Information and Augmented Reality research line of the Navantia-UDC Joint Research Unit and has been funded by it.

**ABSTRACT** The principles of the Industry 4.0 are guiding manufacturing companies towards more automated and computerized factories. Such principles are also applied in shipbuilding, which usually involves numerous complex processes whose automation will improve its efficiency and performance. Navantia, a company that has been building ships for 300 years, is modernizing its shipyards according to the Industry 4.0 principles with the help of the latest technologies. Augmented Reality (AR), which when utilized in an industrial environment is called Industrial AR (IAR), is one of such technologies, since it can be applied in numerous situations in order to provide useful and attractive interfaces that allow shipyard operators to obtain information on their tasks and to interact with certain elements that surround them. This article first reviews the state of the art on IAR applications for shipbuilding and smart manufacturing. Then, the most relevant IAR hardware and software tools are detailed, as well as the main use cases for the application of IAR in a shipyard. Next, it is described Navantia's IAR system, which is based on a fog-computing architecture. Such a system is evaluated when making use of three IAR devices (a smartphone, a tablet and a pair of smart glasses), two AR SDKs (ARToolKit and Vuforia) and multiple IAR markers, with the objective of determining their performance in a shipyard workshop and inside a ship under construction. The results obtained show remarkable performance differences among the different IAR tools and the impact of factors like lighting, pointing out the best combinations of markers, hardware and software to be used depending on the characteristics of the shipyard scenario.

## I. INTRODUCTION

The principles of Industry 4.0 pave the way for the modernization of manufacturing companies, whose automation relies on the application on the latest technologies related to robotics, Industrial Internet of Things (IIoT), Big Data or cyber-physical systems. A lot of research has still to be performed in some of such technologies, which are still not ready for its deployment and intensive use in industrial environments.

One of such technologies to be further studied is Augmented Reality (AR), which has been previously proposed for being used in industrial processes related to the stages of design, commissioning, manufacturing or quality control [1]. In the specific case of Industry 4.0, it can be defined the concept of Industrial Augmented Reality (IAR), which involves the AR hardware and software whose characteristics (e.g., robustness, ruggedness, accessibility, battery life) make them ideal for factories and industrial environments.

This paper analyzes and evaluates the application of different hardware and software IAR tools for their application in the shipbuilding industry. Specifically, this study is part of the Shipyard 4.0 project [2], whose aim is to apply the principles of Industry 4.0 to traditional shipyards to optimize their numerous processes. Shipyard 4.0 is led by Navantia [3] and the University of A Coruña (Spain). Navantia is one of the ten largest shipbuilding companies in the world and has been building military and civil ships since 1717. The Shipyard 4.0 project is divided into different research lines that study the processes that occur in a shipyard. One of such lines is called "Plant Information and Augmented Reality" and its aim is to assess the suitability of IAR tools to provide user-friendly interfaces and information on relevant processes to shipyard operators.

This article presents the IAR architecture developed for the Shipyard 4.0 and shows its application to different use cases. Such an architecture is based on the fog-computing principles [4], being aimed at reducing latency response to provide a realistic AR interaction. After detailing the architecture, its implementation is described for three hardware devices and two AR Software Development Kits (SDKs), which were evaluated in a shipyard workshop and inside a ship.

The present paper includes the following contributions, which, as of writing, have not been found together in the literature:

- It reviews the most relevant research on IAR developments for shipbuilding and smart manufacturing.
- It presents the design of an IAR system that has to operate in an environment as tough in terms of lighting and wireless communications as a shipyard and a ship under construction.
- It evaluates in different scenarios of the shipyard the use of diverse augmented reality hardware and software. In fact, it was not found in the literature any practical evaluation on the application of generic IAR technology in similar use cases and scenarios.

The rest of this paper is structured as follows. Section II reviews the main IAR developments for shipbuilding and smart manufacturing applications, as well as the most relevant IAR hardware and software tools. Section III describes the communications architecture and shows different use cases for the future Shipyard 4.0. Section IV indicates the hardware and software that was selected to implement the IAR solutions. Section V details the experiments performed: their objective, the experimental setup, the results of the different practical tests performed and the key findings. Finally, Section VI is devoted to the conclusions.

## II. RELATED WORK

### A. IAR FOR SMART MANUFACTURING

The progress in electronics, big data, communications, networking, and robotics, together with paradigms like Internet of Things (IoT), enable the development of advanced systems focused on improving energy efficiency [5], [6], automation [7], decision-support [8] or even security [9]–[13]. Today, sectors like transportation [14], Industry 4.0 [15], or even defense [16], [17] are benefiting from such advances. In the case of the AR technologies, although the first pioneering AR developments were presented during the 1960s [18], [19], the field did not gain traction until the industry started to create the first IAR applications in the 1990s. Specifically, the first documented IAR applications were created by Boeing [20] and were aimed at giving step-by-step instructions to factory workers in manufacturing and assembly operations. After these beginnings, interest in IAR started to grow in the industry, but it was not until the late 1990s when the German government pushed AR technology by funding the ARVIKA project [21], [22], whose aim was to develop IAR systems for mobile use. The project created a large IAR consortium that involved companies like Airbus, EADS, BMW, Audi, VW, Daimler, Chrysler or Ford. Several IAR solutions were developed in the ARVIKA project, but most of them stalled at a prototype level, being the most successful, in terms of user acceptance, the AR welding gun developed by the Technical University of Munich and BMW [1].

The progress of IAR systems, although relatively slow, derived into the development of many interesting applications that covered the multiple steps involved in product manufacturing. For instance, IAR systems have been proposed for the early design phases of cars [23] or generic products [24]. In the automotive industry, systems have also been developed for the latest stages of the design, allowing users to select virtual car components to place them on a real car [25].

The most popular IAR applications are for the assembly processes, where it has been recently demonstrated that IAR reduces significantly the number of errors and decreases time and mental workload respect to other approaches [26]. Thus, there exist IAR systems for generic virtual assembly [27], for assembling automotive cockpits [28], for nanomanufacturing [29] or for creating step-by-step instructions for certain assembly processes [30].

IAR applications have also been developed for training operators in different generic manual skills [31] or for performing maintenance in demanding environments like the aerospace sector [32]. Moreover, the automotive industry has shown interest in training, although it seems that such an interest is focused on assembly processes [33], [34].

In addition, IAR has been applied to factory planning [35], to diverse maintenance tasks [36], [37], and to quality controls and accuracy verifications [38], [39].

It has been also studied what an IAR application should include in order to develop a successful solution for manufacturing [1]:

- Reliability. The IAR system should be robust, provide fall-back alternatives and be as accurate as possible.
- User-friendliness. Users have to find the IAR system easy to configure and use. The learning curve should not be steep.
- Scalability. Prototypes have to be created bearing in mind that plant owners and manufacturers have to reproduce and distribute them easily and in large quantities.

Regarding the IAR hardware developments for smart manufacturing, in the last years IAR systems have shown remarkable improvements from the early prototypes of the 1990s [40]. Nowadays, commercial devices are more affordable, are more powerful [41] and introduce numerous benefits that have sparkled again the interest on IAR for smart manufacturing [42]:

- Hands-free viewing of in-situ information.
- It is possible to receive information while carrying out simple tasks, what increases productivity when used properly.
- Real-time user guidance to specific locations, which, in conjunction with a rich context information, helps to find optimal routes.
- Cueing (i.e., use of animations and graphics to indicate the precise point of the item to be manipulated).
- Use of enriched content (e.g., images, 2D diagrams, 3D objects).
- Seamless interaction: operators have only to look at the desired element.
- More immersive experiences. Technology has evolved to the point that it lets users feel immersive first-person experiences that help to provide more emotional engagement in the processes to be performed.

### B. IAR FOR SHIPBUILDING

Most of the IAR smart manufacturing systems previously mentioned can be used directly, or with a light adaptation, for the shipbuilding industry. However, a few researchers have devised IAR solutions that were conceived from scratch for a tough environment like a shipyard or a ship.

Welding is probably the area where more contributions have been proposed [43]–[45]. In [43] it is presented an IAR system that makes use of an augmented welding helmet that displays relevant information about the welding process, including suggestions on possible error corrections. A different approach is detailed in [44], where an IAR interface and a wireless controller are used to allow operators to interact with a welding robot inside a ship. Fast et al. [45] focused on training welders, creating an IAR solution that uses smart glasses and a torch to simulate real-time welding processes.

Another system for training operators is presented in [46]. In such a paper it is described an IAR system that makes use of a pair of AR glasses and a paint gun interface to paint on virtual structures and to see the results immediately, just after finishing the exercise.

Maintenance is also an important topic in shipyards and ships, so some authors have focused their research on it [47], [48]. In [47] it is proposed an IAR system that aims to replace traditional paper and electronic maintenance documents with a tablet that shows step-by-step instructions about the process. In the case of [48], the maintenance tasks, although performed on a military vehicle, are really similar to the ones carried out by mechanics in a ship or in a submarine. Specifically, the system proposes an IAR solution based on a wrist control panel and a Head-Mounted Display (HMD) that allow a mechanic to augment its vision with animations, graphics and text in order to simplify and reduce errors on certain maintenance tasks, which usually take time due to the high component density and the complexity of the maintenance process.

Regarding the use of IAR for design tasks, there are not many academic examples of practical applications. The most relevant paper found in the literature is [49], which describes an IAR system that allows users to detect discrepancies between the ship CAD design models and the actual construction. Thus, an operator can visualize the pipe construction data in a ship and then perform changes to adapt the CAD design to the reality of the construction (for instance, due to tolerances, it is typical to find misalignments or collisions between mounting elements).

Finally, it is worth mentioning that there are several companies that offer IAR solutions that have been developed explicitly for shipbuilding. For example, Newport News Shipbuilding [50] is developing IAR safety, training, operation and maintenance applications aimed at increasing savings in different shipbuilding processes. Index AR Solutions [51] develops IAR applications for construction, operations, maintenance and utilities for different industries, including shipbuilding. The other relevant company is BAE Systems [52], which has already used IAR interfaces for building offshore patrol vessels and to design its Type 26 frigate.

### C. IAR HARDWARE AND SOFTWARE

In the last years, different companies have presented and commercialized AR hardware interfaces that can be used in industrial environments [53]–[65]. Most of such interfaces make use of smart glasses [53]–[59], [63]–[65], but there are also helmets [60], [61] and other HMD devices [62].

The price of the previously cited IAR hardware devices currently ranges from US $ 500 to US $ 5,000, so they are still too expensive for their massive use. The price tag is high due to the amount of hardware that is embedded in a really compact and light device. In fact, most of the devices carry similar hardware to the one embedded into a smartphone: between 1 and 2 GB of RAM, up to 128 GB of ROM, a multi-core processor, Wi-Fi/Bluetooth connectivity, one or two cameras, an Inertial Measuring Unit (IMU) and multiple sensors (e.g., barometer, luminosity sensor, magnetometer, GPS). In relation to the camera, which is key in IAR applications, some devices include a deep focus camera [55], while others improve accuracy by adding thermal and infrared cameras [60]. In terms of battery life, most IAR devices allow for reaching four hours of use, although some do not (at least with an active use) [61], [64].

It can be stated that, as of writing, is has not been found the perfect IAR wearable device, which should include the following characteristics [41]:

- The field of view should be as wide as possible. 30°(horizontally) are usually the minimum recommended for providing a good user experience.
- The IAR interface should be as light as possible, since it has to be worn during the whole day.
- Ideally, the batteries should last through the working day.
- Optical and retinal projection should be used, since video-based display technologies incur in delays that

harm the user experience.

- Voice-based interaction is recommended to enable hands-free operation, although voice processing has still to be improved to work properly in noisy industrial environments.

With respect to AR software for developing industrial applications, fortunately, it is a step ahead of hardware in terms of progress, existing a wide variety of AR SDKs and libraries available, like Augmenta Studio [66], AugmentedPro [67], ALVAR [68], ARLab [69], ARmedia [70], ARToolkit [71], ArUco [72], Aurasma [73], BaZar [74], BeyondAR [75], Beyond Reality Face [76], Catchoom [77], IN2AR [78], Instant Reality [79], Layar [80], Mixare [81], OpenSpace3D [82], SSTT [83], UART [84], Vuforia [85], Wikitude [86] or ZappWorks [87]. Some of them are open-source [71], [72], [74], [82], [84], other are free [68]–[70], [73], [75], [77]–[81], [85], [86], and others offer commercial versions [66], [68]–[70], [73], [76]–[80], [83], [85]–[87].

This software implements all or part of the three main features ideally required by an IAR application: fast processing for overlapping virtual elements in the field of view, implementation of recognition and tracking algorithms to detect and follow elements, and speech and/or gesture recognition mechanisms. Not all these features are necessary in every IAR application, but the choice of the right software should take into account the application environment, the development platform and the hardware (processing power, sensors) required by the application.

For the sake of clarity, Table 1 summarizes the main IAR hardware and software solutions mentioned in this Section.

TABLE 1: IAR hardware and software.

| **Hardware & software** | **Solutions** |
|---|---|
| Hardware | Atheer Air glasses [53]<br>CHIPSIP Sime glasses [54]<br>Epson Moverio BT-200/300/2000 [55]<br>ODG R-7 smartglasses [56]<br>RECON Jet Glasses [57] Sony<br>Smarteye [58]<br>Vuzix [59] DAQRI<br>Smarthelmet [60] Microsoft<br>HoloLens [61] Fujitsu<br>Ubiquitousware HMD [62] Laster<br>WAV9 [63] Optinvent<br>ORA-2 [64] Penny C Wear<br>30 Extended [65] ARToolkit [71] |
| Open-source software | ArUco [72]<br>BaZar [74]<br>OpenSpace3D [82]<br>UART [84]<br>ALVAR [68] |
| Free software | ARLab [69]<br>ARmedia [70]<br>ARToolkit [71]<br>Aurasma [73]<br>BeyondAR [75]<br>Catchoom [77]<br>IN2AR [78]<br>Instant Reality [79]<br>Layar [80]<br>Mixare [81]<br>SSTT [83]<br>Vuforia [85]<br>Wikitude [86] |

## III. IAR SYSTEM DESIGN

### A. SHIPYARD'S IAR USE CASES

The project Shipyard 4.0 aims to study the application of IAR to all the processes involved in shipbuilding. Specifically, after analyzing the state-of-the-art and the processes carried out in workshops and in a ship, the following use cases were selected as the most promising in terms of potential efficiency improvements achieved through the application of IAR:

#### 1) Plant information

shipyard operators usually rely on paper to identify assets (e.g, pipes, machines, pallets) and determine which action should be performed according to the work orders. An IAR application can suppress the vast majority of the paperwork by providing dynamic real-time information about the assets. For instance, in Figure 1 it is shown the evaluation of an IAR application based on Vuforia that displays information about the characteristics of a pipe on a smartphone. In this case, the IAR marker acts as a unique identifier associated with the identification number of the pipe in the information system so that the IAR application can show contextual information like material, size and destination of each individual pipe.

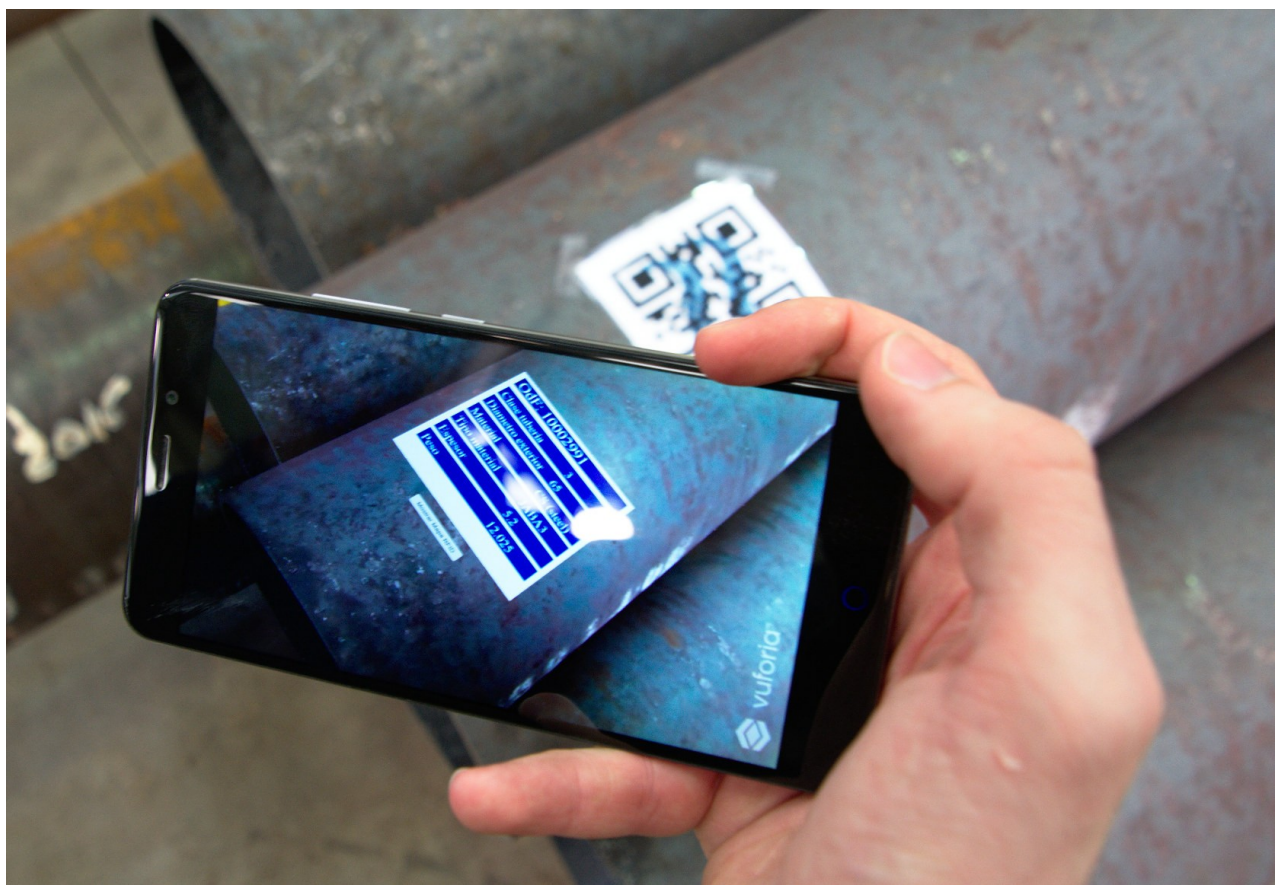

FIGURE 1: Pipe information on a smartphone.

#### 2) Quality control

after performing almost every process in a shipyard workshop, a quality check needs to be carried out. An IAR application can help by superimposing the 3D CAD model created by Navantia's designers on the real piece in order to determine visually the differences. Moreover, during the quality control it is possible to automatically detect deviations from the 3D CAD model and point them out to the operator. Such a control requires first to scan the item by using 3D cameras and then to make use of reconstruction software to generate the new model.

#### 3) Assistance in the manufacturing process

an IAR application can guide operators by indicating step-by-step instructions. This can be performed by showing 3D models on tangible interfaces located on the workbench, where a marker may act as a spatial reference.

#### 4) Asset location

a workshop is a large environment where assets can be anywhere. IAR solutions can help to locate them by pointing at the specific place or area where an asset is. For instance, in Navantia's pipe workshop it is being deployed a pipe location system based on active UHF RFID tags [88], [89] that can interact with an IAR application in order to show such locations in portable devices like tablets or IAR smart glasses, as illustrated in Figure 2. UHF RFID was chosen among other technologies (BLE, WiFi, ultrasounds, UWB, ZigBee, Z-Wave, WirelessHART, RuBee) after considering factors such as deployment, presence of metals, presence of water, exposure to liquids, acids, salinity, fuel or other corrosive substances, potential communications interference, reading distances, tolerance to high temperatures, pressure, battery duration, mobility and cost.

#### 5) Visualization of installations

in a ship it is usual that part of the infrastructure (i.e., piping, wiring) is installed behind bulkheads, roofs or ceilings, which makes its location difficult. IAR can overlap the 3D design to reality and then show such a location. This is specially useful for maintenance and fault repairs. Moreover, it is also interesting to monitor the shipyard infrastructure, which can even be linked to the IIoT data to show relevant notifications and variables in real time. As an example, in Figure 3 it is shown the monitoring view of Navantia's shipyard in Fene (Galicia) when displayed through Microsoft HoloLens glasses in an office.

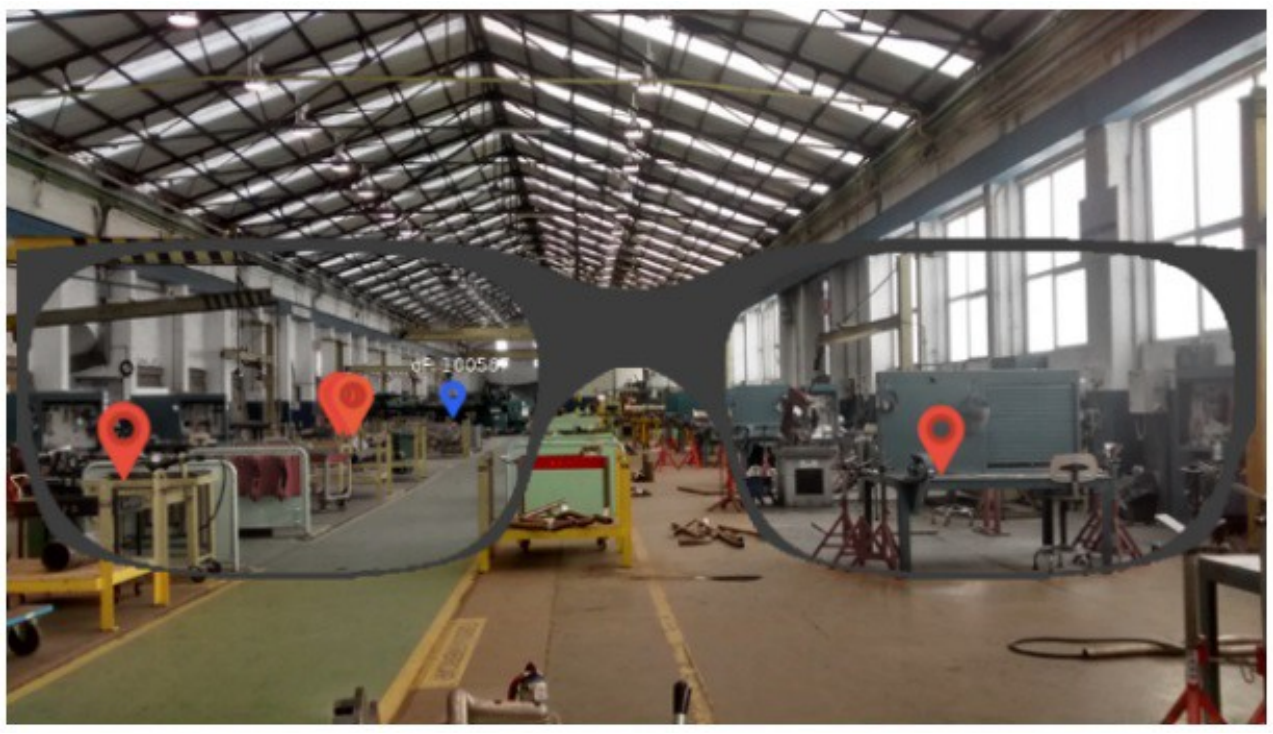

FIGURE 2: Localization of pipes using both IAR and an RFID-based system.

#### 6) Warehouse management

since shipyard warehouses are actually quite large, it is helpful to provide operators with an IAR-based guidance system that allows them to locate and store items faster and to decrease collection and storing errors. Moreover, an IAR application might show the content of the different shelves when looking for specific parts (illustrated in Figure 4).

#### 7) Predictive maintenance

The data from quality controls and sensors installed throughout the different workshops can be processed and shown dynamically to operators in order to detect anomalies and identify with precision possible current or future problems.

#### 8) Augmented collaboration and reporting

IAR application can enhance collaboration in real time with the objective of providing remote guidance on the resolution of incidents or to clarify visually certain events. This is possible thanks to be ability of IAR applications for sharing the Point of View (POV) of the operator and enabling the superposition of information over the actual images seen by the operator. Moreover, it is possible to record video and audio notes for enriching reports and, thus, provide better clarifications than traditional text and image-based reports.

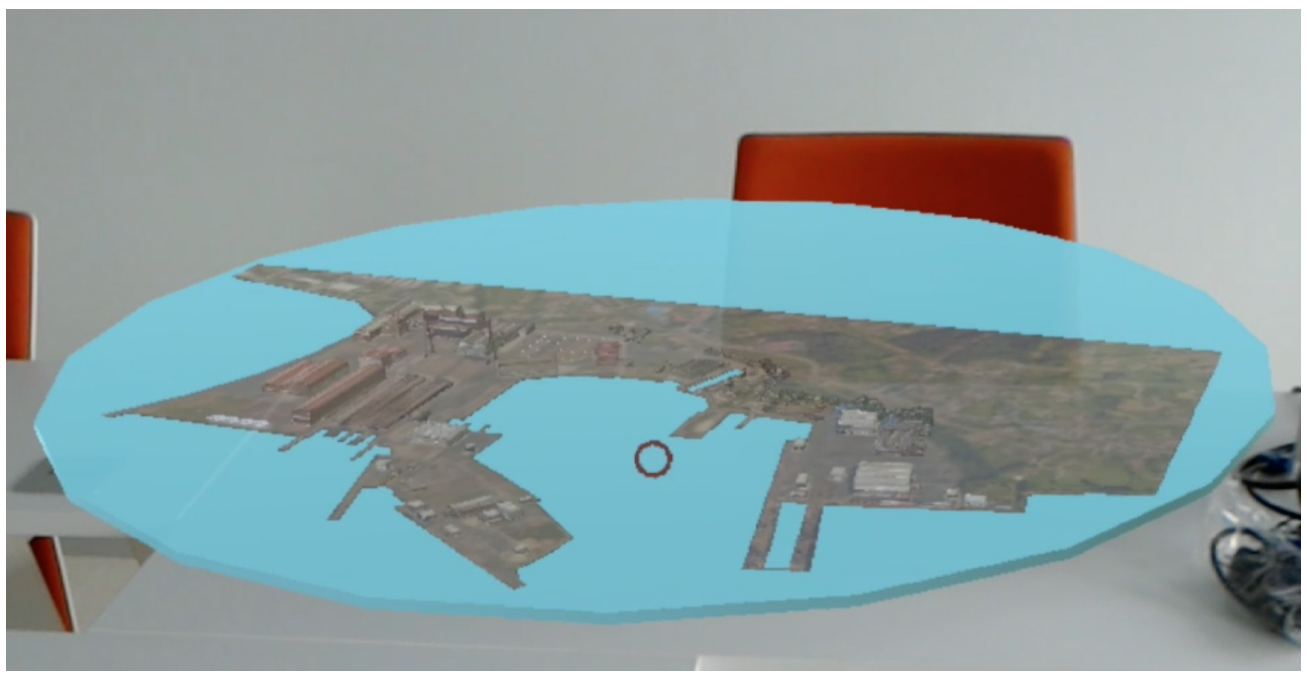

FIGURE 3: Shipyard model through HoloLens.

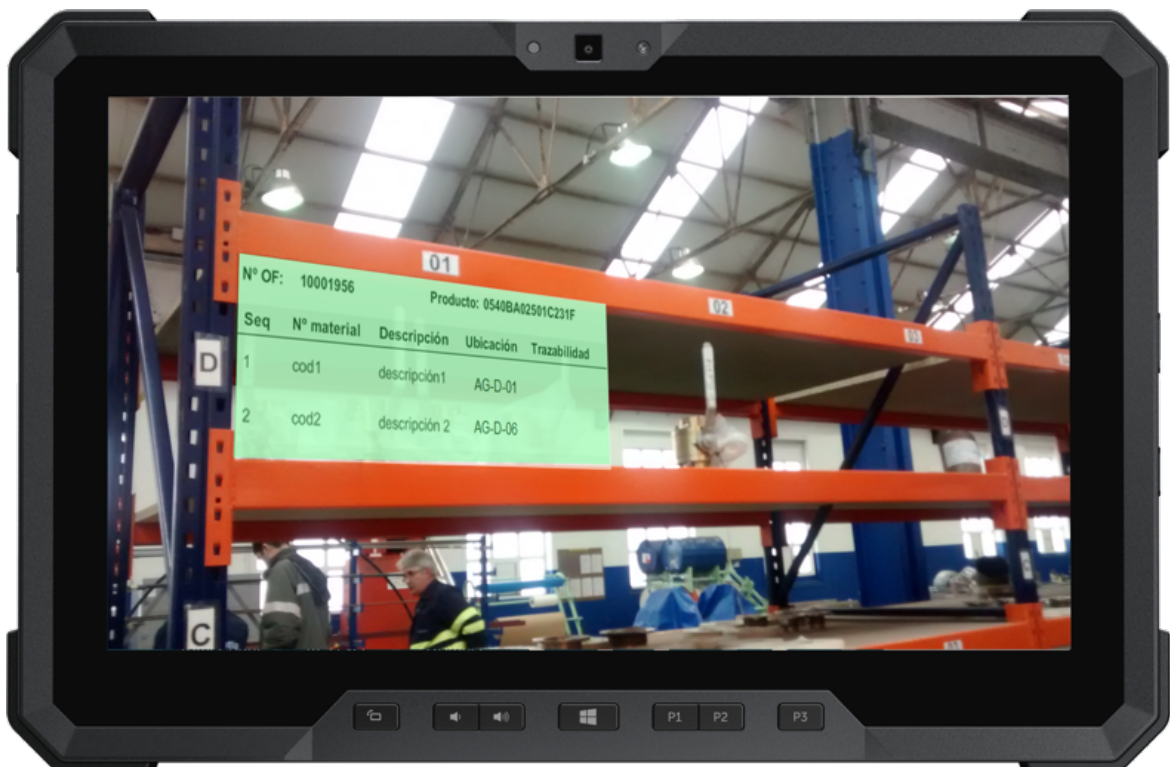

FIGURE 4: Content of one of the shelves of a warehouse.

#### 9) Block assembly

a ship is made out of different elements that are first assembled into blocks and then such blocks are assembled together. The correct alignment of each piece of the hull is critical, especially the exact positioning of the pipes whose misalignment can cause severe problems during the assembly process. IAR applications can help operators by guiding them through the various processes and simulating the joints by projecting the augmented 3D model of a block next to a real one.

### B. COMMUNICATIONS ARCHITECTURE

In order to develop the applications related to the use cases described previously, the proposed IAR system makes use of the fog computing communications architecture depicted in Figure 5. Fog computing extends cloud computing by moving part of the computational and communication capabilities of the cloud close to the sensor nodes [4]. Such a move provides an IAR system with remarkable benefits:

- Local fog nodes allow for minimizing latency response. This is required in dynamic real-time IAR systems that usually suffer from high lags when accessing data stored in the cloud or in remote servers.
- The fog is able to distribute computational power and storage among local gateways. Such resources are accessed through services, which are especially helpful when operators wear IAR devices that are usually resource-constraint on purpose to preserve lightness and to extend battery life. Therefore, such IAR devices can delegate certain demanding tasks or large file storage to the fog.
- Resource distribution also improves mobility and location awareness, providing services to mobile and location constrained users.
- The fog is able to connect IAR devices in different physical locations of the shipyard, thus easing collaboration among operators.
- The fog is usually composed by cheap and small gateways, which make the system highly flexible and easy to scale.

Therefore, fog computing is ideal for providing IAR services thanks to its ability to support physically distributed, low-latency and QoS-aware applications that decrease the network traffic and the computational load of traditional cloud computing systems. Navantia's IAR fog-computing architecture is composed by three layers:

- **Node layer**: it includes all the IAR devices that interact with the services provided by the fog layer. The fog layer also exchanges data with sensor networks that conform the shipyard IIoT ecosystem, and RFID readers that make use of a pipe positioning system [88].
- **Fog layer**: it consists of one or several single-board computers (SBCs) that are installed in fixed positions throughout the shipyard workshops and in a ship. Each SBC acts as a gateway and provides fog services. In the case of the AR service, it supplies IAR devices with localized data and responds faster than the cloud, thus acting as a proxy caching server for IAR data.
- **The cloud**: this is where the data is stored when received from the multiple sources of the shipyard. It is also the place where Navantia runs its own compute-intensive services and the ones offered through third-party software (i.e., SAP for Enterprise Resource Planning (ERP), FORAN for shipbuilding, Windchill for Product Lifecycle Management (PLM), and ThingWorx as IoT platform).

It is worth mentioning that the communications between each IAR device and the fog layer are performed through Wi-Fi connections, since Navantia has already deployed IEEE 802.11 b/g/n infrastructure. However, note that communications inside a ship suppose a challenge for electro-magnetic propagation due to the presence of numerous large metal elements. As of writing, Power Line Communication (PLC) is being evaluated in Navantia's ships to provide connectivity to IAR devices, but further research should be performed because, although the system works most of the time, its network speed is influenced by the electrical interference coming from electric circular saws and other tools that demand high-current peaks. Therefore, it is important to emphasize that the results presented in this paper do not evaluate the communications performance of the architecture, but the marker recognition and tracking performance of the IAR devices.

## IV. IMPLEMENTATION

### A. SELECTED HARDWARE

In order to implement the node layer of the communications architecture described in Section III-B, three devices were tested: a mid-range smartphone (UMI Super), a rugged tablet (Panasonic FZ-A2mk1) and a pair of smart glasses (Epson Moverio BT-2000).

The UMI Super is an Android 6.0 smartphone with a 64-bit 2.0 GHz octa-core processor (Helio P10, MTK6755), a 5.5-inch screen, 4 GB of RAM, 32 GB of ROM and a 13 MP rear camera. The Panasonic FZ-A2mk1 is a rugged tablet that also runs Android 6.0 and that embeds an Intel AtomTM x5-Z8550 processor (it reaches up to 2.4 GHz), an Intel HD 400 graphics card, a 10.1-inch screen, 4 GB of RAM, 32 GB of internal memory and a 8 MP rear camera with autofocus and LED light. Regarding the EPSON Moverio BT-2000 smart glasses, they run Android 4.04 and their hardware, whose internal architecture is depicted in Figure 6, is less powerful than the one used by the other two selected devices: the glasses include a Texas Instruments OMAP 4460 1.2 GHz dual-core processor, 1 GB of RAM, 8 GB of internal memory (although it can be expanded to up to 32 GB through a microSDHC card), a 0.42-inch 24-bit TFT display with a FOV of 23°, and a 5 MP camera.

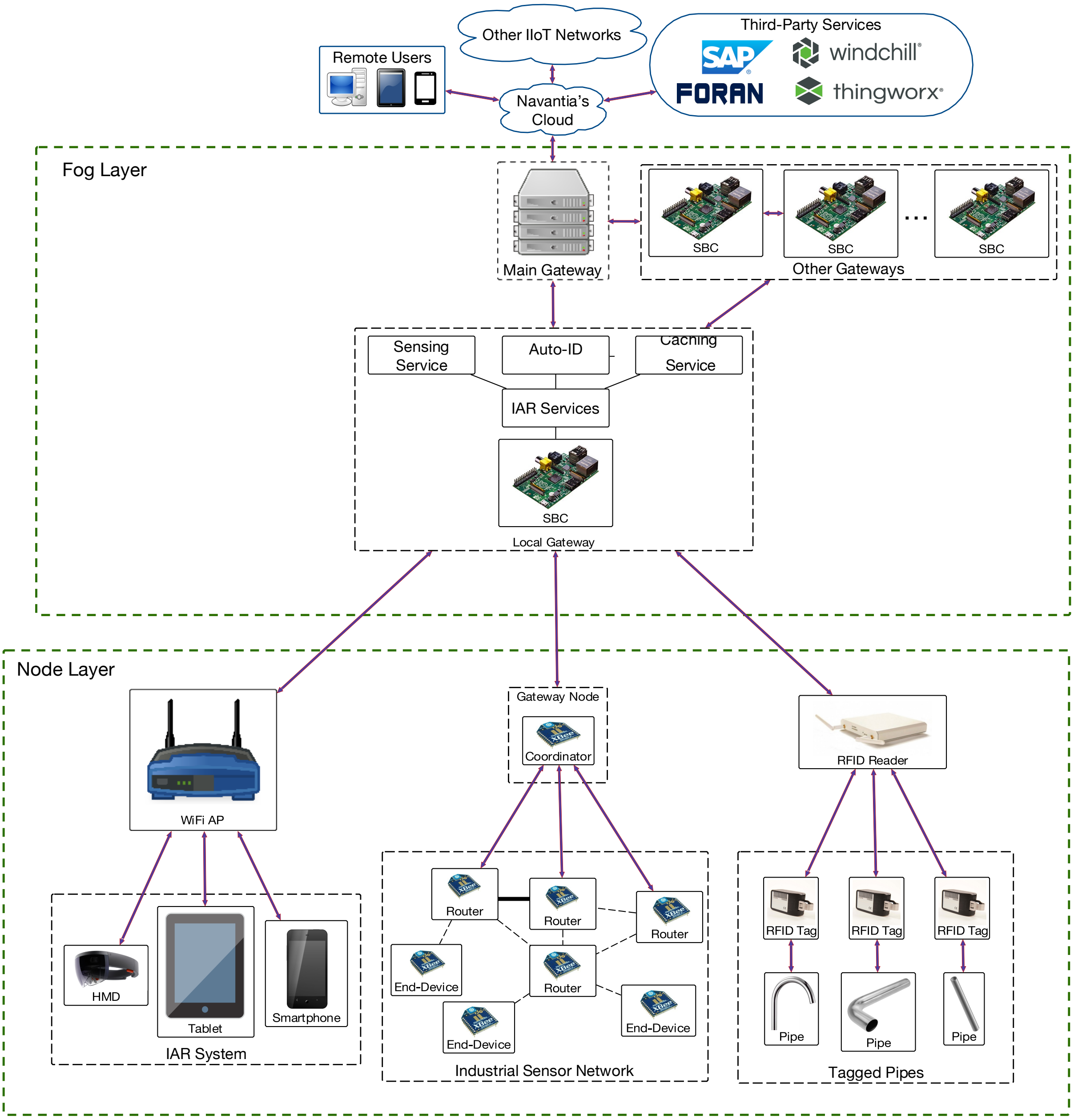

FIGURE 5: Architecture of the IAR system proposed.

Note that user interaction differs among the selected hardware. In both the smartphone and the tablet the interaction is performed through a touch interface and the device must be pointed at the AR marker to visualize the contextual augmented information. In contrast, the Epson Moverio BT-2000 glasses have a see-through display on which the augmented information is projected directly, so users usually do not have to perform any actions to recognize the markers. In addition, the glasses can make use of gesture, touch and voice recognition.

Another approach would consist in utilizing Spatial Augmented Reality (SAR) and use projectors to display graphical information onto physical objects, thus detaching the technology from the operators and allowing seamless collaboration among workers. However, the large extension of Navantia's shipyard in Ferrol (1,000,000 $m^2$) makes it really difficult and expensive to deploy an effective system to cover most of the production areas, whereas an HMD display can be carried by the operator all over the shipyard. In addition, recent improvements in HMDs and see-through displays allows them to offer appealing features that generate an immersive personalized user experience that cannot be currently obtained through SAR.

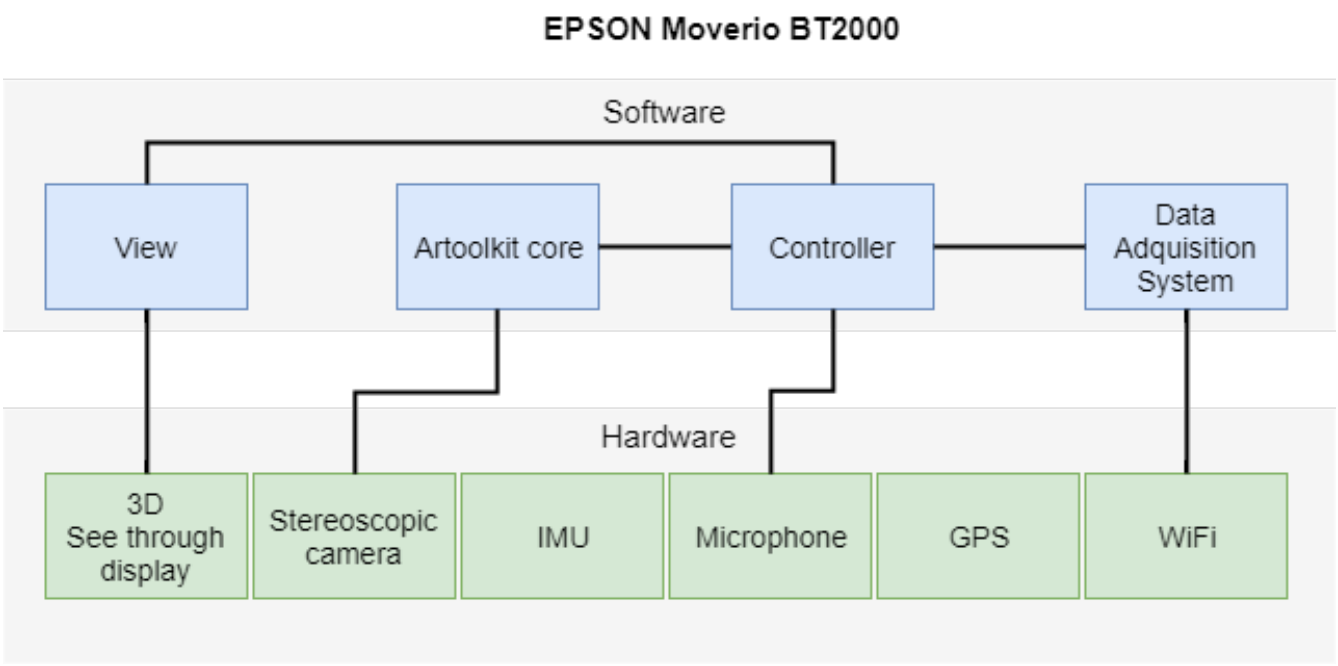

FIGURE 6: EPSON Moverio BT-2000 software and hardware architecture.

### B. SELECTED SOFTWARE

After analyzing the advantages and disadvantages of the software mentioned in Section II-C, two SDKs were chosen as the most promising for developing IAR applications with the hardware selected: Vuforia and ARToolkit. Their main characteristics are compared in Table 2.

Vuforia has support for Unity 3D [90] and ThingWorx (an IoT platform that is currently being tested by Navantia), and is one of the most popular AR SDKs at the moment. Companies like Sony, Toyota, Adidas and BMW are supporting its development and using it. It enables to create IAR systems on the most common development platforms, but it is not possible to use it with certain smart AR glasses, like the Epson Moverio BT-2000.

Vuforia uses a cloud service to extract features or patterns from images previously uploaded. The sharper and more complex an image, the better for pattern recognition. In Vuforia, QR markers can be generated easily, having different characteristics that make them suitable for Vuforia's recognition mechanism. This is the reason why QR markers were one of the candidates chosen for the experiments performed in Navantia's shipyard (described later in Section V).

Regarding ARToolkit, it also has a large community of developers and users, and, since it is open-source software, it makes it possible to modify it and create a version compatible with virtually any ad-hoc device. Moreover, ARToolkit includes algorithms optimized for long-distance pattern recognition, which is a desirable feature in multiple shipbuilding scenarios. In addition, it is worth noting that this SDK implements a specific marker recognition algorithm based on image features that has been optimized for a built-in type of maker called binary marker that can be read really fast, at long distances, and that can include error detection and correction codes to make them more reliable.

## V. EXPERIMENTS

### A. OBJECTIVE

Several experiments were performed in order to determine the feasibility of deploying practical IAR applications in an industrial environment as harsh as a shipyard and a ship in terms of luminosity and visual interference. Specifically, the aim of the experiments was to determine the recognition distances obtained by each IAR system when varying three parameters: the type of marker, the lighting conditions and the reading angle.

First, note that, in an IAR system, the luminosity of the environment, the type of light and the light temperature are essential for implementing reliable and fast IAR applications. Since AR algorithms make use of computer vision techniques based in thresholding and color measurement metrics, the lighting characteristics can impact the system performance significantly. In addition, electrical interference produced by the industrial machinery may affect sensor readings and their accuracy.

Regarding the reading angle of the markers, in some experiments the tests were performed at different angles in order to determine the maximum angle at which the markers were detected. Note that, in real-world scenarios, shadows and reflections can appear on the marker, what impacts the recognition distance and the marker detection success rate.

In addition, the material and the type of marker, as well as their size, influence the detection range of the system. Regarding the marker material, when a marker is printed in a laser printer with regular matte paper, the toner usually causes a glossy finish that can affect the recognition distance under certain lighting circumstances. However, the impact of the reflections on the reading distances varies depending on the scenario.

Finally, it is worth mentioning that, although the experiments evaluated IAR frameworks that rely on physical markers, it is possible to make use of markerless recognition systems (e.g., Microsoft HoloLens), which may avoid part of the problems previously described. Such markerless systems are still a minority in the world of IAR and, although they work fine in certain environments, they currently have some limitations that reduce their applicability in a shipyard:

- Due to the limited memory of current markerless IAR hardware, its use tends to be limited to rooms or reduced areas, which have to be mapped (usually on-the-fly) by using different sensors. It is possible to load data dynamically, as the user enters a new area, but that makes the development more complex and requires fast storage and processing hardware. In contrast, marker-based systems work like simple hash tables that associate an identifier with certain content or event, although they can also implement sophisticated multi-marker positioning algorithms.
- Most markerless IAR systems are based on detecting physical characteristics of an object or a place. Such a detection is usually more complex, what makes algorithms more compute intensive and, therefore, more powerful hardware is required. Moreover, the detection of physical patterns is not only more complicated computationally, but it is also influenced by ambient light, being often even more sensitive to changes in the environment than other marker-based detection techniques.

TABLE 2: Comparison between ARToolKit and Vuforia.

| Product | License type | Tracking | Detection Techniques | Description |
|---|---|---|---|---|
| ARToolkit | Open Source, Commercial SDK | Android, iOS, Linux OSX, Windows | Marker, NFT | Recently acquired by Daqri. Compatible with Epson Moverio and Daqri Smarthelmet. |
| Vuforia | Free, Commercial SDK | Android, iOS | Markers, NFT, Visual Search | It incorporates computer vision technology to recognize and track planar images and 3D objects in real time. This image registration capability enables developers to position and orient virtual objects in relation to real world images when they are viewed through the camera of a device. Moreover, Vuforia's SDK provides a list of 100,000 commonly used English words that can be incorporated into Text Recognition apps. It is compatible with different Epson Moverio smart glasses, ODG R-7 and Microsoft Hololens. |

- There are also markerless systems that complement visual pattern recognition with embedded sensor measurements. A problem may occur in a shipyard when a markerless system relies on a GPS receiver or in Wi-Fi signals to locate a user or a room, since in certain environments (for instance, indoors or inside a ship under construction) such positioning techniques may not work properly.
- Some markerless systems rely on incremental tracking, combining information collected from different sensors and from the camera. Nonetheless, a dynamic environment like a shipyard workshop or a ship that is being built, where the geometrical structure of the place changes through time, can mislead the markerless IAR system.

### B. EXPERIMENTAL SETUP

The tests presented in this article evaluate the performance of the three IAR hardware devices described in Section IV-A, in different scenarios and with varied markers:

- UMI Super and Panasonic FZ-A2mk1: since they run the same operating system (Android 6.0), the same Vuforia and ARToolkit-based applications could be tested on both. A screenshot of the ARToolkit application is shown in Figure 7. The Vuforia application, although different in its inner workings, has a really similar interface (in Figure 8).
- Epson Moverio BT-2000: these glasses required a specific development optimized for their hardware. However, its visual interface is basically the same as the one designed for the ARToolKit smartphone/tablet application, as it can be observed in Figure 9, where it is shown during one of the lab tests.

In both Vuforia and ARToolKit applications it is assumed that the markers are associated with one of the key products produced in a shipyard, a pipe, whose main characteristics are displayed on the IAR device. Specifically, in Figures 7 to 9 it is shown the work order (*OdF*), the type of pipe (*Clase*), the diameter (*Diámetro*), the material, the type of material (*Tipo*

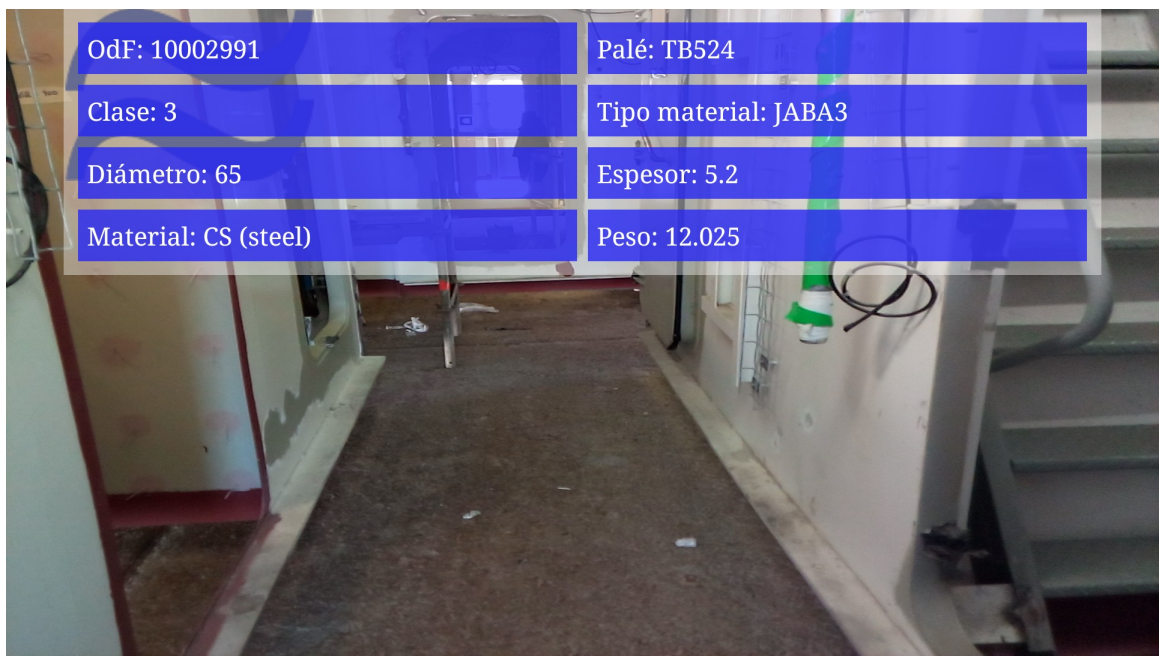

FIGURE 7: Screenshot of the ARToolkit application.

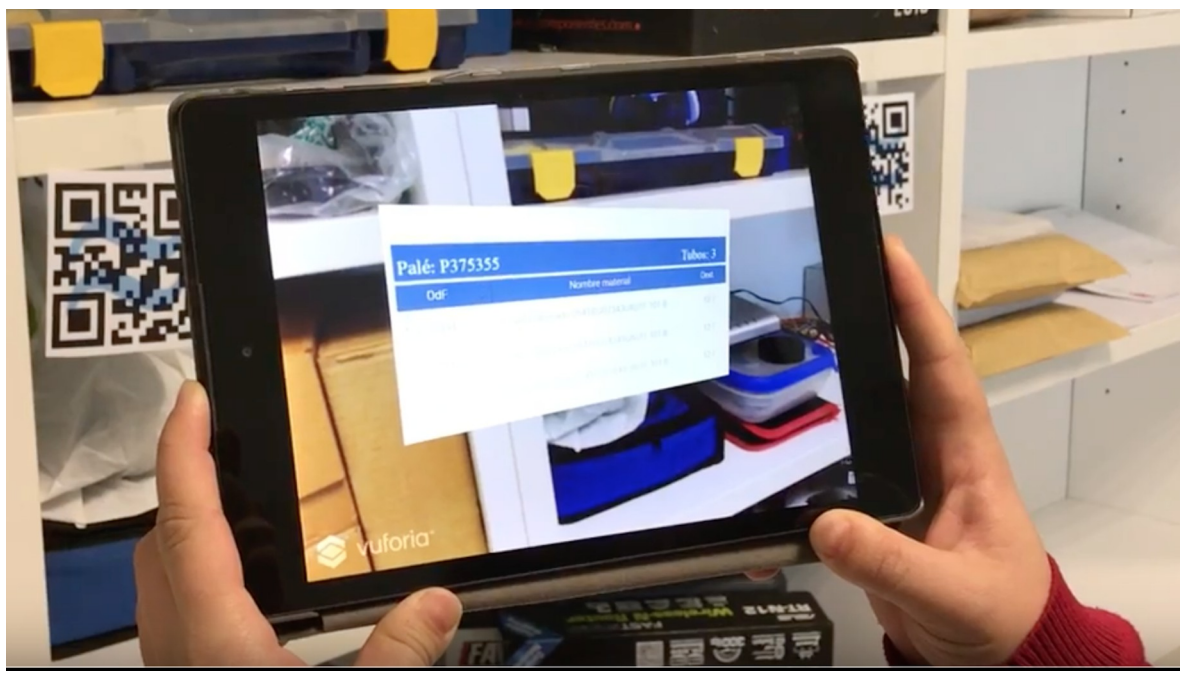

FIGURE 8: Screenshot of the smartphone/tablet Vuforia application.

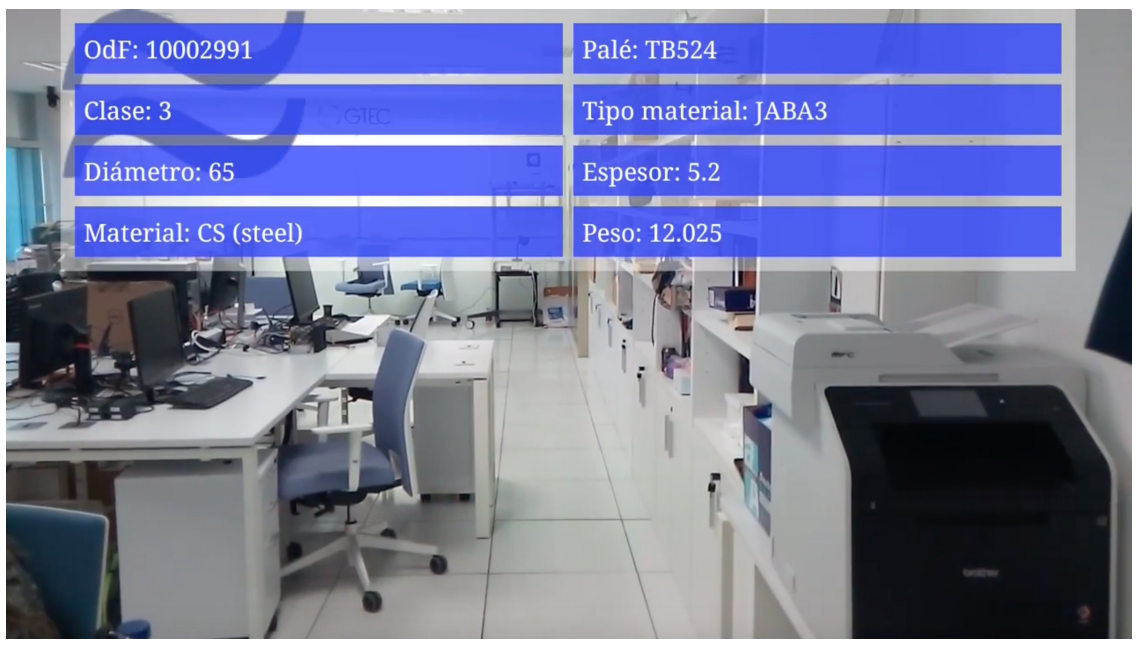

FIGURE 9: Screenshot of the Epson Moverio BT-2000 ARToolkit application.

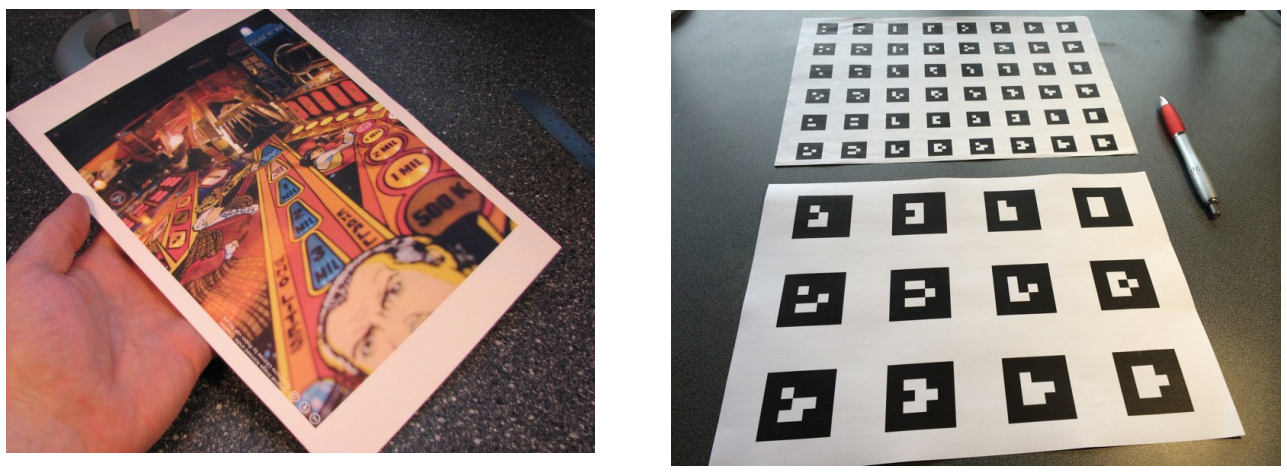

(a) (b)

FIGURE 10: Examples of NFT marker (a) and 2D barcode binary markers (b).

*material*), the pallet where the pipe is located (*Palé*), and the pipe thickness (*Espesor*) and weight (*Peso*).

When using the ARToolKit and Vuforia applications, two basic types of markers were detected (in Figure 10), which require making use of different recognition techniques:

- **Natural Feature Tracking (NFT)**: it consists in storing and detecting recognizable patterns present in markers. It is a method supported by both ARToolkit and Vuforia that makes it possible to detect a marker even when it is partially hidden or when luminosity is low.
- **Binary code detection**: it is a method that is only available in ARToolkit. It is optimized for faster recognition with specific markers associated with binary codes.

The following markers were the ones selected for the experiments (they are shown in Table 3 together with their main characteristics):

- **Custom square**: this marker was designed under the specifications of ARToolkit, which indicate that a marker must include a square with a black frame and then contain a simple drawing inside. Nonetheless, it is also possible to invert the colors (the frame and drawing would be white, while the background would be black). In any case, this is a black and white marker that relies on contrast differences to be detected. The simpler the drawing inside, the easier it is detected, but less marker combinations are possible.
- **Square QR marker**: it was designed following the specifications of Vuforia, which establish the goal of obtaining the highest possible number of reference points in an image. For instance, QR codes are good candidates due to their shape and properties. The cloud system of Vuforia is in charge of extracting the marker characteristics and provides a file that contains the binary data used to recognize the marker. In the experiments performed, in order to use this kind of markers with ARToolkit, it was used the NFT detection technique.
- **Rectangular QR marker**: it is like a square QR marker, but larger, in order to include more information and ease the recognition. This increase in size improves Vuforia's detection, but, due to its shape, it is really difficult to detect it with ARToolkit.
- **Binary markers with or without BCH**: they are designed for ARToolkit and optimized for being recognized easily. The ARToolkit software allows for generating this type of markers with an area ranging from three to six square bits. The higher the number of bits, the more the number of possible makers, although such markers are also larger. Specifically, the number of possible markers available depends on the number of rows and columns in the marker and on the error detection and correction algorithms enabled. The use of better error detection/correction algorithms results in a smaller set of possible markers, but they will lower the probability of being misrecognized. In general, it is better to use markers with the largest possible Hamming distance, as this results in the lowest likelihood of one marker being misrecognized as a different marker. Thus, for the sake of fairness, during the tests performed two types of binary markers were selected: one without error detection/correction codes and another one with a BCH (13,9,3).

Finally, it is worth mentioning that the selected markers were printed into different materials and some were laminated to protect them from water and dust. Those makers would be used in areas exposed to rain and dirt, while non-laminated markers would be placed indoors, in less aggressive scenarios, like certain pipe storage locations.

After testing such markers in the lab, very similar maximum detection and tracking distances were obtained, so, for the sake of brevity, the results shown in the next Section are the ones referred to laser printed paper markers that were not laminated.

### C. WORKSHOP TESTS

The IAR system was first tested in the pre-assembly P1 workshop, a large building located in the facilities of the shipyard that Navantia owns in Ferrol (Spain). Figure 11 shows one of the moments during tests in the workshop, when detecting a marker successfully at a long distance.

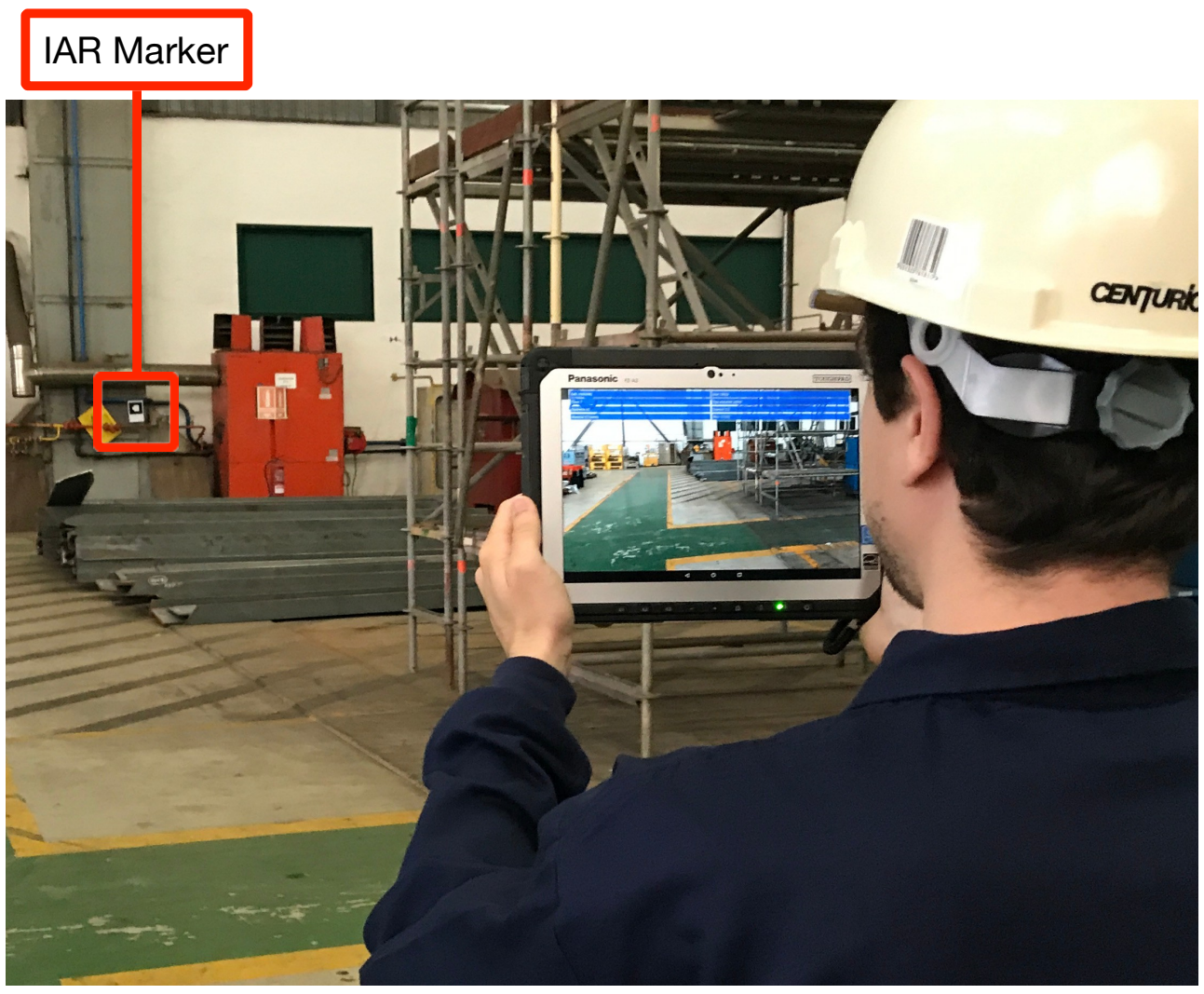

FIGURE 11: Long distance reading test in the pre-assembly P1 workshop.

TABLE 3: Type of markers used in the experiments.

| Type | Size (mm) | Supported SDK | Example | Distance | Tolerance to Partial Concealment | Max. Number of Markers |
|---|---|---|---|---|---|---|
| Custom square | 80x80 | ARToolkit, Vuforia | 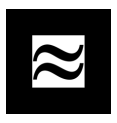 | High | Very low | High (>1000) |
| Square QR marker | 80x80 | ARToolkit, Vuforia | | Low | Medium | High (>1000) |
| Rectangular QR marker | 80x400 | Vuforia | | Medium | High | High (>1000) |
| Binary markers A/C | 190x190 | ARToolkit | 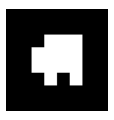 | High | Very low | 8192 |
| Binary marker B (BCH (13,9,3)) | 190x190 | ARToolkit | | High | Very low | 512 |

It is important to emphasize that in this environment luminosity is low due to the height of the ceilings, where the spotlights are installed. Nevertheless, the natural lighting projected through the main workshop entrance and the sky-lights increase the luminosity, although it varies throughout the day. Due to such a variability in the lighting conditions, exhaustive measurements of the lighting level were carried out. Differences during the day were remarkable as the sun rose. For instance, early in the morning, natural lighting was scarce, so values under 160 lx were obtained. However, luminosity values increase gradually until noon, reaching around 280 lx. Therefore, light changes influenced the experiments because the luminosity was not stable.

Table 4 compares the maximum reading distances obtained for different marker types and sizes, for the three hardware devices selected, when running the ARToolkit and Vuforia applications developed.

In the case of Vuforia, two distances were obtained: the maximum at which a marker was recognized and the one at which the marker was tracked. The latter is related to how far the AR device can be moved away after recognizing a marker. This difference exists because, by design, once Vuforia detects a marker, it is possible to track it at a longer distance without losing its position within the image. In ARToolkit, since detection and tracking distances are very similar, they have not been specified in the Table.

The values contained in Table 4 allow for observing that, when lighting conditions were around 220 lx, ARToolkit reached recognition distances between 6 and 10 m with the optimized binary markers. As light conditions improved, the recognition distances for the same markers went up to 26 m. This distance makes it possible to stand in the middle of the workshop and detect properly 190 mm-wide markers located on the walls. The custom square marker also achieved very good recognition distances (almost 7 m) with ARToolkit despite its small size (80×80 mm). Vuforia also recognized correctly the custom square marker, but recognition distances were clearly shorter (less than 2 m). In addition, it can be observed that Vuforia does not detect the binary markers optimized for ARToolkit, since it uses natural feature points for their recognition and these type of simple markers present just a few characteristic points.

Vuforia gets its best results with QR markers, although most distances are lower than 2 m. On the contrary, AR-Toolkit has difficulties in detecting QR markers because of the necessity of having to use of the NFT technique. Nonetheless, when ARToolkit detects a QR marker, in most cases the recognition distance is similar to the one obtained by Vuforia.

Regarding the Rectangular QR, it is not correctly recognized by ARToolkit due to its shape, while Vuforia obtains its best results with it due to the increase in size and, therefore, in the number of characteristic points.

With respect to the hardware, both the tablet and the smartphone obtain similar results, while the Epson Moverio smart glasses are not able to reach detection distances as long as the other IAR devices.

After analyzing all the results, it can be concluded that the best recognition distances are obtained with the binary markers and ARToolkit, although it is important to note that such distances will decrease remarkably in low light scenarios. It is also worth mentioning that the differences in recognition distances among the three binary markers are very small. The longest recognition distance is achieved with the Panasonic FZ-A2mk1 tablet and ARToolkit when detecting the binary marker C with 280 lx of light (26 m).

In Table 5, it is represented the detection rate for each marker, which is calculated as the number of times that the initial detection is correct with respect to the total number of times that is viewed through an IAR device at a specific distance. The three different IAR devices of the previous tests were evaluated under the same lighting conditions, averaging the results for 15 executions of every test and when reading the markers at a 0°angle (i.e., in parallel to them). It is also important to note that the two markers at the top look the same, but the second one includes error correction/detection. At the view of the results, it can be observed that the markers that make use of error correction codes perform better, although above 15 m, one of the markers showed a remarkable decrease in its detection rate.

TABLE 4: Maximum marker recognition and tracking distances obtained at the pre-assembly workshop.

| Marker | Type | Average Luminosity (lx) | Recognition/Tracking distance (m) | | | | |
|---|---|---|---|---|---|---|---|
| | | | UMI Super | | Panasonic FZ-A2mk1 | | Epson Moverio |
| | | | ARToolkit | Vuforia | ARToolkit | Vuforia | ARToolkit |
| 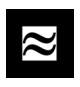 | Custom square | 160 | – | 0.19/1.20 | 1.84 | 0.19/1.39 | – |
| | | 220 | 2.46 | 0.21/1.70 | 2.23 | 0.21/1.70 | 2.31 |
| | | 280 | 6.50 | 0.30/1.70 | 6.70 | 0.39/0.98 | 5.70 |
| 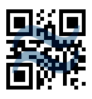 | Square QR marker | 160 | – | 0.32 | 0.30 | 0.40/1.31 | 0.35 |
| | | 220 | 0.38 | 0.40/1.40 | 0.48 | 0.50/1.40 | 0.40 |
| | | 280 | 0.60 | 0.50/2.00 | 0.56 | 0.60/1.94 | 0.67 |
|  | Rectangular QR marker | 160 | – | 0.68/1.62 | – | 0.34/1.50 | – |
| | | 220 | – | 0.80/1.58 | – | 0.40/2.00 | – |
| | | 280 | – | 1.30/3.00 | – | 1.29/2.80 | – |
| 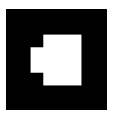 | Binary marker A | 160 | – | – | 2.36 | – | – |
| | | 220 | 6.40 | – | 9.60 | – | 7.8 |
| | | 280 | 25.00 | – | 26.00 | – | 12.50 |
| 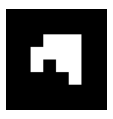 | Binary marker B (BCH (13,9,3)) | 160 | – | | 2.17 | – | 9.4 |
| | | 220 | 6.32 | | 7.90 | – | 9.4 |
| | | 280 | 22.30 | | 25.00 | – | 11.24 |
| | Binary marker C | 160 | – | | 2.40 | – | 5.80 |
| | | 220 | 6.00 | | 9.50 | – | 9.20 |
| | | 280 | 25.80 | – | 26.00 | – | 12.60 |

TABLE 5: Detection rate of the binary markers.

| Marker | Error Correction | Average Detection Rate | | |
|---|---|---|---|---|
| | | 10 m | 15 m | 18 m |
| | None | 100% | 98% | 90% |
| | BCH (13,9,3) | 100% | 100% | 99% |
| | BCH (13,9,3) | 100% | 80% | 60% |
| | None | 100% | 50% | 15% |

Finally, it should be mentioned that the two markers at the top obtain better results due to the fact that they are less complex than the others, so they can be detected easier, even when they are partially hidden.

### D. EVALUATION OF THE IAR SYSTEMS IN A SHIP

The IAR system was also tested in a modular offshore patrol vessel that was under construction in the shipyard (in Figure 12). In Figure 13 it is depicted the blueprint of the ship, which includes two red circles that indicate the two areas where the tests were performed: the one at the bottom is the dining room and the one at the top is the bridge.

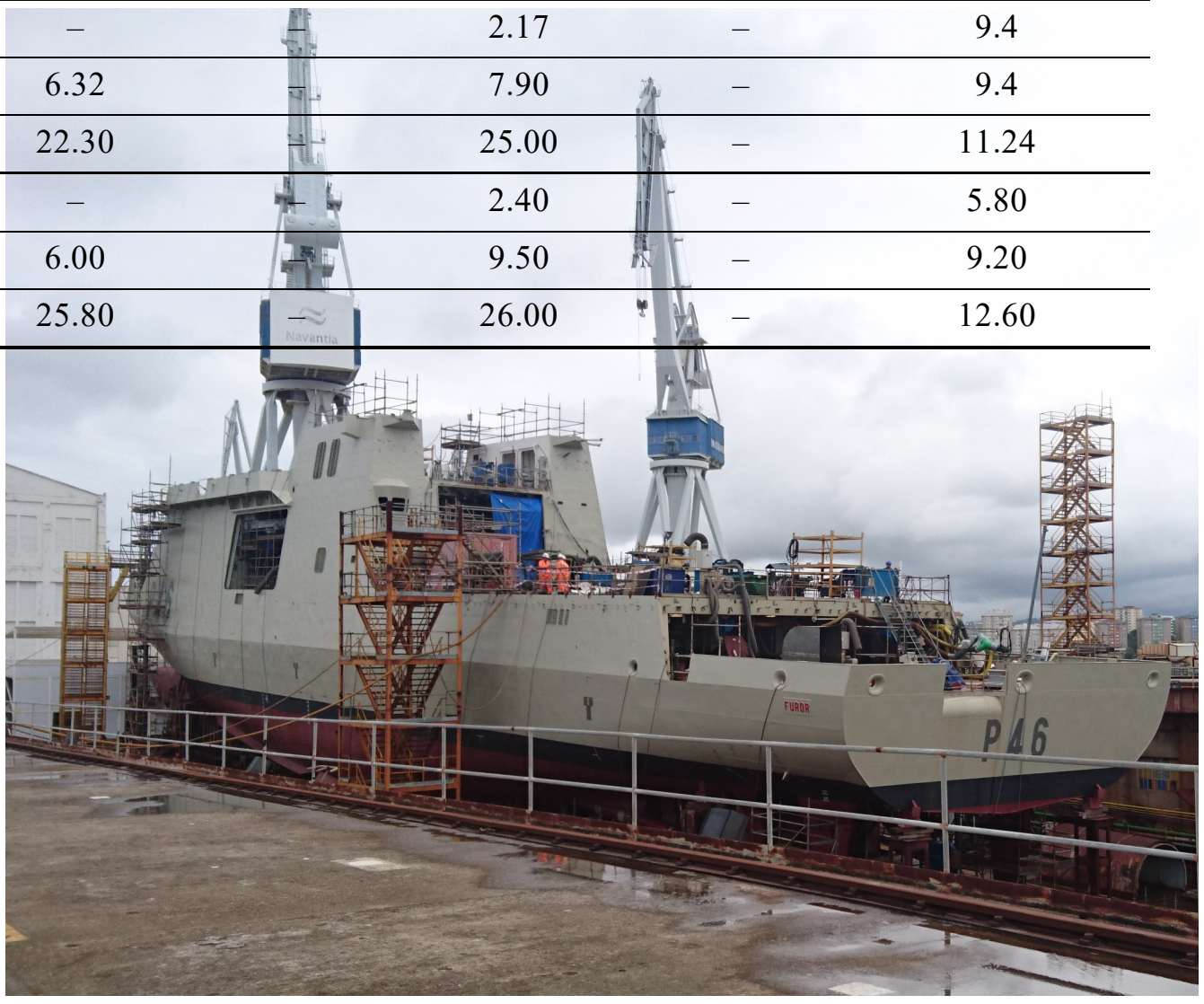

FIGURE 12: Modular offshore patrol vessel where the IAR tests were performed.

In the dining room there is no natural light and all the light comes from fluorescent lamps installed temporarily on the ceilings and walls for the construction works. In addition to the non-homogeneous lighting, the color temperature is quite warm, what reduces the pure white levels of the markers and also greatly affects the thresholding calculations.

The light measurements performed in the dining room averaged 160 lx, existing certain places with better luminosity thanks to their direct exposure to the light of a fluorescent lamp. To carry out the experiments, three different locations with different lighting levels were selected in the dining room. Figure 14 shows scenarios A and B, both of which present very low levels of illumination (as an example, Figure 15 shows one of the moments during the tests with such a lighting). The worst lighting conditions occur in scenario A, where the fluorescent lights located next to the marker were turned off and there was light coming from the back of the camera. In contrast, indirect light was received at Scenario B from several fluorescent lights in the room, and there was no light behind it, what allowed the camera to adjust the sensitivity of the sensor properly. Regarding Scenario C, it is shown in Figure 16. In this position markers receive direct light from one of the fluorescent lamps.

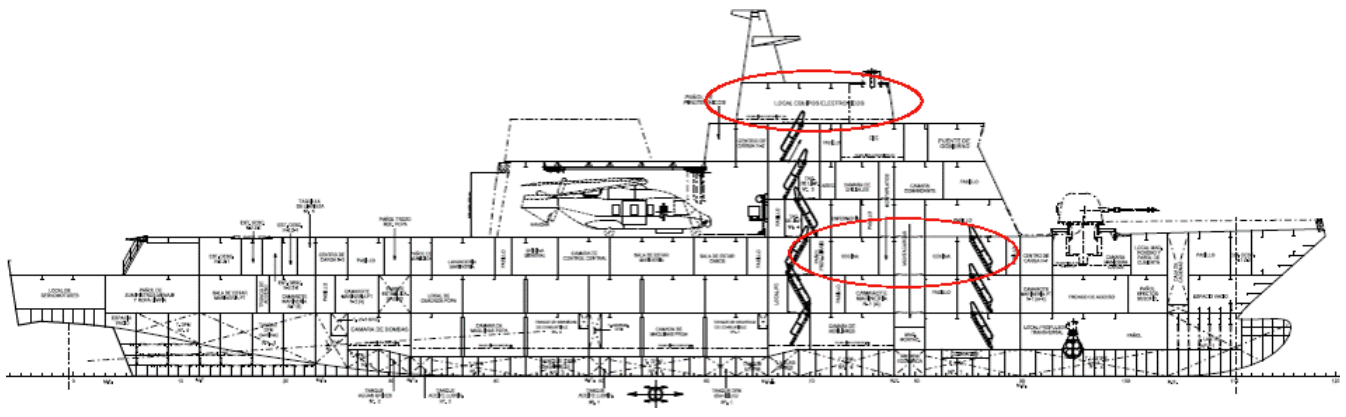

FIGURE 13: Blueprint of the ship.

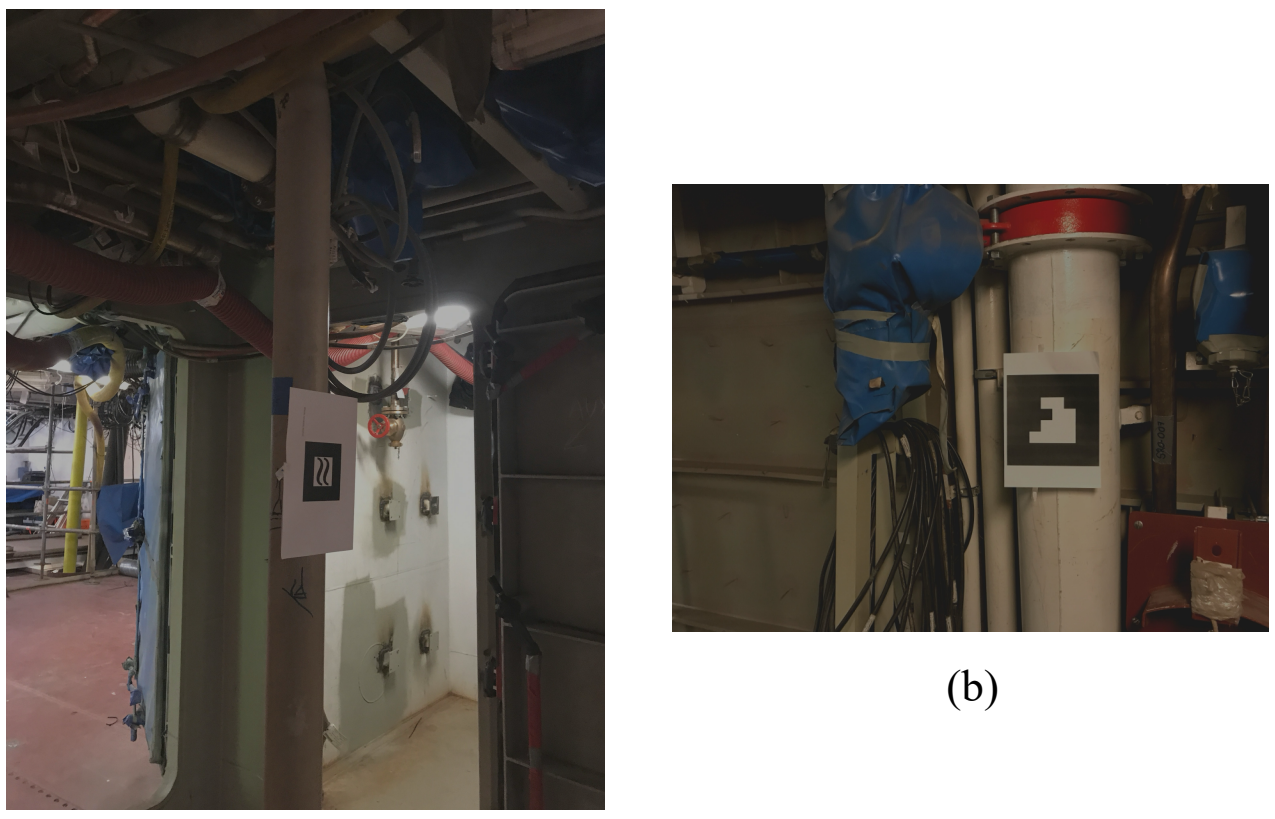

(a) (b)

FIGURE 14: a) Scenario A and b) Scenario B.

Table 6 shows the maximum recognition and tracking distances obtained for the three scenarios. At the view of the results, the first thing to notice is that, due to the lack of light in the room, the majority of the tests failed. Moreover, in these scenarios, in general, recognition and tracking distances are decreased with respect to the ones obtained in the workshop. This is basically caused by the fact that the white pixels in the markers are shown darker, what misleads the recognition algorithms.

After analyzing the results for the three scenarios, it can be concluded that the hardware characteristics of the camera of an IAR device have a significant influence on the recognition performance of the system. The UMI Super smartphone is the device that performs the worst due to the ISO sensitivity of its camera. On the contrary, the Epson Moverio BT-2000 camera adjusts sensitivity quite well to the lighting conditions, while the tablet camera performance is somewhere between the other two devices.

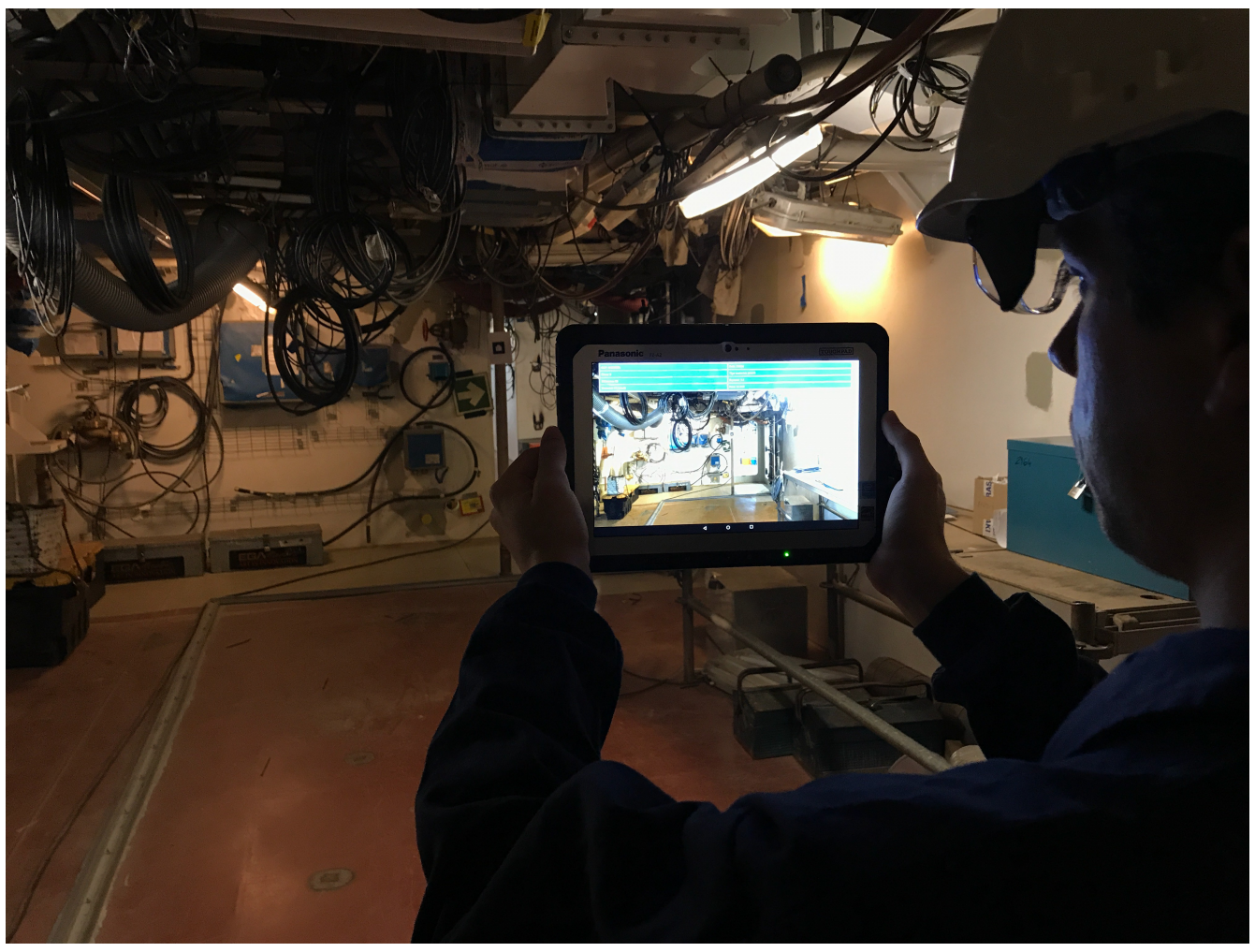

FIGURE 15: Reading tests in the ship under low lighting.

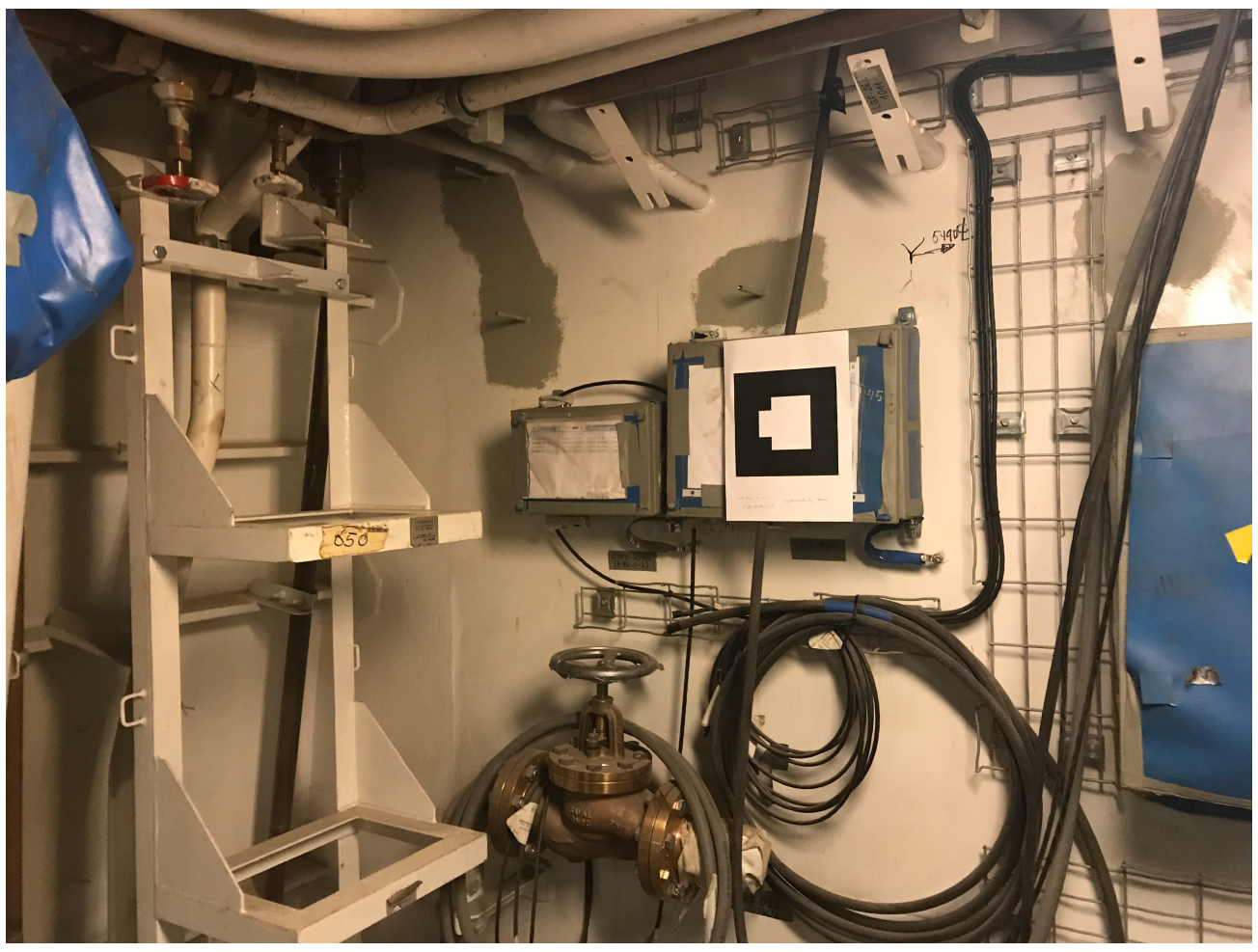

FIGURE 16: Scenario C.

Another important factor is the type of marker recognition system. As it can be observed in Table 6, Vuforia is less affected by the lighting conditions than ARToolkit, being able to recognize patterns with poor lighting. However, the recognition distances obtained with Vuforia are quite short. In contrast, ARToolkit recognition system fails more under low light conditions, especially with custom square and binary markers, due to the image processing method implemented by the SDK for the recognition of this type of markers. However, the longest recognition distances were obtained by the Epson Moverio glasses and the tablet for such types of markers, reaching more than 10 m. Among the different markers, the square QR marker, which is detected by using the NFT technique, is read slightly better than the others under low illumination.

TABLE 6: Recognition distances in the dining room of the ship.

| Marker | Type | Scenario | Recognition/Tracking distance (m) | | | | |
|---|---|---|---|---|---|---|---|
| | | | UMI Super | | Panasonic FZ- A2mk1 | | Epson Moverio |
| | | | ARToolkit | Vuforia | ARToolkit | Vuforia | ARToolkit |
| 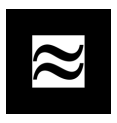 | Custom square | A | – | – | – | – | – |
| | | B | – | – | 3.70 | – | 3.30 |
| | | C | – | – | 4.00 | – | 3.70 |
| | Square QR marker | A | – | 0.38/1.45 | – | 0.39/1.20 | – |
| | | B | 0.59 | 0.40/1.55 | 0.42 | 0.43/1.40 | 0.80 |
| | | C | 0.64 | 0.44/1.50 | 0.50 | 0.45/1.56 | 0.87 |
| | Rectangular QR marker | A | – | 1.07/2.20 | – | 1.00/2.10 | – |
| | | B | – | 1.36/2.70 | – | 1.29/2.05 | – |
| | | C | – | 1.40/2.60 | – | 1.40/2.30 | – |
| 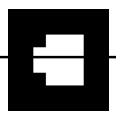 | Binary marker | A | – | – | – | – | – |
| | | B | – | – | – | – | 3.80 |
| | | C | – | – | 9.56 | – | 13.40 |
| 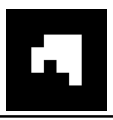 | Binary BCH(13,9,3) | A | – | – | – | – | – |
| | | B | – | – | 2.40 | – | 3.80 |
| | | C | – | – | 9.00 | – | 10.90 |
| 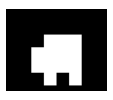 | Binary marker 4×4 | A | – | – | – | – | – |
| | | B | – | – | 3.90 | – | 3.80 |
| | | C | – | – | 10.20 | – | 11.00 |

The second set of experiments in the ship was performed in the bridge. In this scenario, the lighting conditions are quite different from the dining room, because in the bridge there are windows through which sunlight enters the room. Nonetheless, sunlight causes back-light and reflections that may impact the recognition performance of the IAR systems. Due to the similarity in performance between the tablet and the smartphone in this scenario, for the sake of brevity, only the results for the smartphone will be shown.

The results of the tests performed in this area can be seen in Table 7. The first noticeable change in this scenario is that, many more markers can be recognized in comparison with the dining room due to the ambient light. However, since light reflections occur because of the position of the windows, for the sake of fairness, measurements had to be taken at different angles. Thus, Table 7 shows the maximum recognition distances at different angles (0° (in parallel to the IAR camera), 30°, 45° and 60°). It can be noticed that marker detection distances at 0° are reasonable for the lighting conditions of the scenario, but, in general, as the reading angle increases, light reflections reduce remarkably the recognition distance.

An example of the influence of the light on the measurements is shown in Figure 17, where it can be seen the light coming from a window behind the marker. This environment makes it difficult for the IAR system to detect the marker properly, since the camera has to adjust the sensitivity to match the light of the scene. With this level of ambient light, ARToolkit obtains better results in all the test cases, even with markers created explicitly for Vuforia like the square QR marker. However, as it was previously mentioned, the rectangular QR marker is not read with the ARToolKit application (except in a very specific angle) due to its shape. It can also be observed that, in this scenario, the UMI Super smartphone and the Epson Moverio glasses perform in a similar way. These results mean that the technical characteristics of the hardware become more important when the light is low, as it is more difficult for the camera to adapt contrast levels. On the contrary, when lighting is good, the differences between different hardware devices disappear.

### *E. KEY FINDINGS*

The previously detailed experiments allow for drawing different conclusions regarding the performance of the software and hardware chosen in the scenarios selected:

- In general, ARToolKit obtains longer maximum recognition and tracking distances than Vuforia. Specifically, ARToolKit obtains its best results with binary markers, reaching in some scenarios more than 25 m. Furthermore, binary marker detection rates increase when using BCH error correction codes.
- The recognition distance does not exceed 2 m when using NFT techniques. In contrast, in the tests performed with the ARToolkit's binary marker detection algorithm, it was possible to perform the recognition fast at more than 20 m, although the markers had to be fully visible and there were good lighting conditions.

TABLE 7: Maximum recognition/tracking distances in the bridge.

| **Marker** | **Type** | **Reading Angle** | **Recognition/Tracking distance** (m) | | |
|---|---|---|---|---|---|
| | | | UMI Super | | Epson Moverio |
| | | | ARToolkit | Vuforia | ARToolkit |
| | Custom square | 0 | 2.62 | 0.38/1.80 | 5.51 |
| | | 30 | 3.91 | 0.19/1.41 | 3.72 |
| | | 45 | 4.57 | – | 3.27 |
| | | 60 | 3.24 | – | 1.59 |
| | Square QR marker | 0 | 0.70 | 0.47/2.29 | 0.76 |
| | | 30 | 0.54 | 0.44/1.90 | 0.43 |
| | | 45 | 0.36 | 0.32/1.43 | – |
| | | 60 | – | – | – |
| | Rectangular QR marker | 0 | – | 1.25/3.00 | – |
| | | 30 | – | 0.72/2.70 | 0.43 |
| | | 45 | – | 0.43/2.12 | – |
| | | 60 | – | – | – |
| | Binary marker | 0 | 6.59 | – | 4.56 |
| | | 30 | 5.74 | – | 1.94 |
| | | 45 | 2.47 | – | – |
| | | 60 | 1.51 | – | – |
| | Binary BCH (13,9,3) | 0 | 6.75 | – | 6.00 |
| | | 30 | 2.91 | – | 1.25 |
| | | 45 | 0.30 | – | 0.80 |
| | | 60 | – | – | – |
| | Binary marker | 0 | 12.43 | – | 13.24 |
| | | 30 | 8.00 | – | 13.24 |
| | | 45 | 4.31 | – | 3.72 |
| | | 60 | – | – | – |

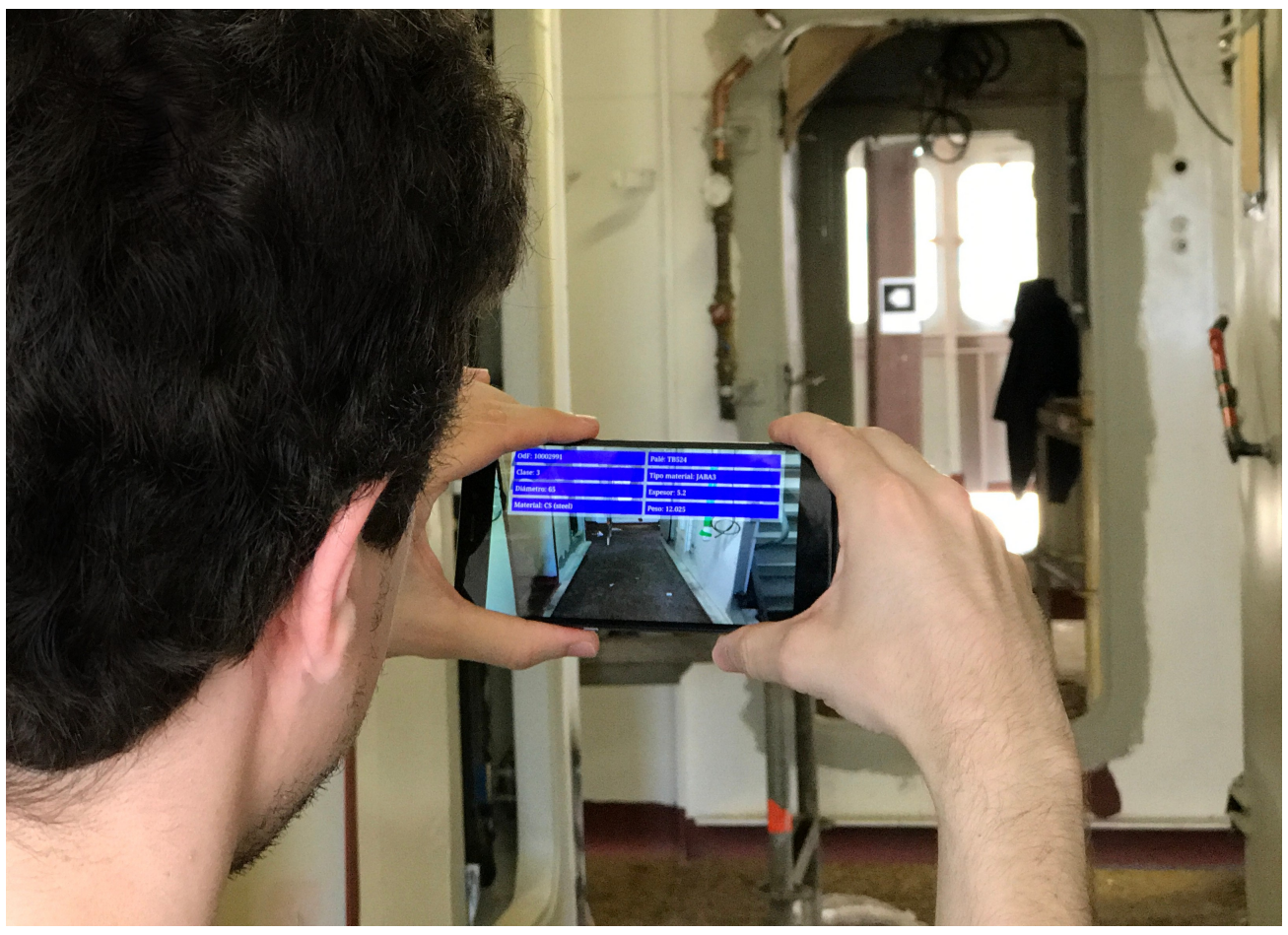

FIGURE 17: One of the IAR tests in the bridge.

- Lighting conditions impact remarkably the performance of marker-based IAR systems. In fact, most of the systems tested become useless under low lighting because of the misbehavior of their recognition algorithms. In such conditions, Vuforia is less affected by lighting than ARToolkit. However, with a regular luminosity, reflections might become an issue and, as the reading angle increases, they reduce the recognition distance.
- With regard to the hardware, it can be stated that, when luminosity is low, camera features are the most important factor to obtain a good performance in an IAR system. In particular, the dynamic range of the sensor has a remarkable impact in the recognition rate. This is due to that fact that, in order to detect the marker edges correctly, the contrast should be high in any lighting condition. In this regard, the UMI Super smartphone has a low range 13 MP camera, which causes the performance to deteriorate very noticeably when the lighting conditions are not good or when there is backlighting. The Panasonic tablet and the Epson glasses have a camera with a similar dynamic range, wider than the smartphone, so they obtain better performance in low-luminosity conditions. On the contrary, when lighting is good, the differences among the hardware devices disappear and the software algorithms are more important.

- A possible solution for the illumination issues could consist in adding specific lamps to the markers so that they have their own light source. Moreover, it would be also possible to attach some type of directional torch in the operator helmet in order to make the light point towards the same position the IAR device camera is facing. These two approaches would undoubtedly improve the performance of the system, although it would be necessary to study their feasibility in detail.
- Other camera characteristics are not so relevant in terms of IAR performance. For instance, the resolution of the camera is not a limiting factor, since the size of the frames should be restricted to a maximum of 1080 pixels in high to obtain an acceptable performance.
- The material and quality of the marker barely influence the recognition distance. The different tests performed with laser-printed, ink-printed and laminated markers showed that the non-laminated markers produced less reflections, but such effect just decreases slightly the maximum tracking/detection distance. However, note that the quality and durability of laminated markers make them more robust when they are exposed to dust, water or grease.
- Finally, the vast majority of current IAR hardware devices can be still considered experimental developments, what makes it difficult their integration with existing IAR frameworks and the implementation of new features. An open-source framework makes it possible to develop adaptations to work with different platforms and opens the door to future optimizations of the final system, something unfeasible in the case of proprietary software. This open-source feature is what allowed us to adapt ARToolkit to work on the Epson Moverio glasses during the development stage for the experiments, but it was not possible to do it with a proprietary solution like Vuforia.

## VI. CONCLUSIONS

This article described the selection and validation of the necessary technologies to design an IAR system for developing applications for an Industry 4.0 shipyard. It was first conducted a detailed review on IAR solutions for smart manufacturing and shipbuilding, and on the IAR hardware and software that might be used in such fields. After such a review, it was described the main IAR use cases related to shipbuilding and the IAR communications architecture deployed by Navantia in its shipyard in Ferrol (Spain). Next, it was selected the IAR hardware and software for the experiments, which, through multiple tests, confirmed the impact of lighting and the reading angle in current hardware and software IAR solutions. Moreover, it was corroborated the importance of the recognition and tracking algorithms, and that ARToolKit reading distances were far longer than the ones obtained by Vuforia under normal lighting conditions. In addition, it was concluded that ARToolKit, since it is open-source, is a better long-term choice than other proprietary solutions when developing applications for IAR devices because of the possibility of extending the software.

To summarize, IAR can bring multiple benefits to an Industry 4.0 shipyard, but current marker-based IAR solutions, although they work fine in many scenarios, still have to progress and be adapted to low lighting situations. Markerless systems seem promising, but systems like Hololens, Meta2 or the ARCore platform require further study to make them fulfill the specific requirements and demands of a Shipyard 4.0.

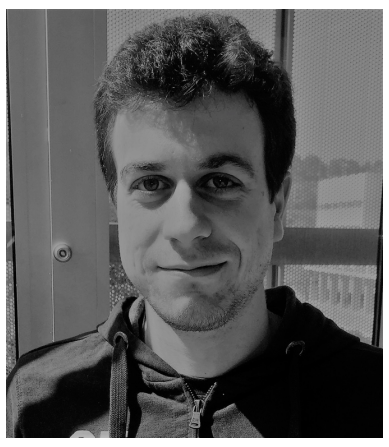

ÓSCAR BLANCO-NOVOA received his B.Sc. in Computer Science in 2016 with mention in computer engineering and Information Technology at the University of A Coruña (UDC). During the last years in college combined his studies with a job as a software engineer at a private company. Currently he is studying his Master's degree in Computer Science and works at the Group of Electronic Technology and Communications (GTEC) in the Department of Computer Engineering (UDC). His current research interests include Energy Control Smart Systems, Augmented Reality and Industry 4.0.

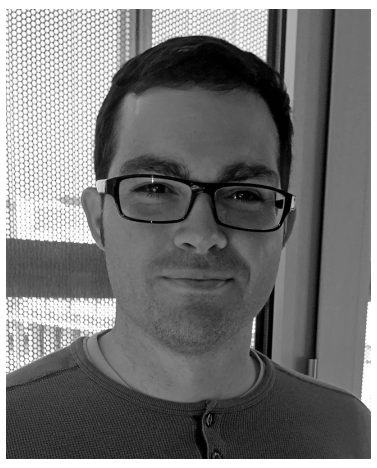

TIAGO M. FERNÁNDEZ-CARAMÉS (S'08-M'12-SM'15) received his MSc degree and PhD degrees in Computer Engineering in 2005 and 2011 from University of A Coruña, Spain. Since 2005 he has been working with the Department of Computer Engineering at the University of A Coruña. His current research interests include IoT systems, RFID, wireless sensor networks, augmented reality, embedded systems and wireless communications.

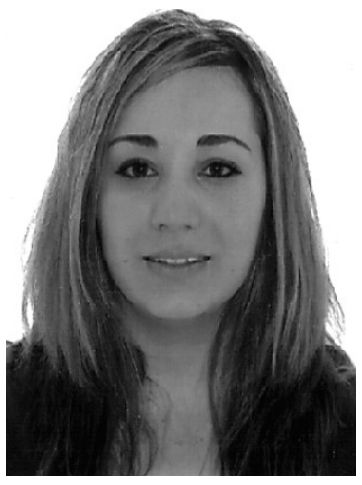

PAULA FRAGA-LAMAS (M'17) received the M.Sc. degree in Computer Engineering in 2008 from University of A Coruña (UDC) and the M.Sc. and Ph.D. degrees in the joint program Mobile Network Information and Communication Technologies from five Spanish universities: University of the Basque Country, University of Cantabria, University of Zaragoza, University of Oviedo and University of A Coruña, in 2011 and 2017, respectively. Since 2009, she has been working with the Group of Electronic Technology and Communications (GTEC) in the Department of Computer Engineering (UDC). She is co-author of more than twenty peer-reviewed indexed journals, international conferences and book chapters. Her current research interests include wireless communications in mission-critical scenarios, Industry 4.0, Internet of Things (IoT), Augmented Reality (AR), RFID and Cyber-Physical systems (CPS). She has also been participating in more than twenty research projects funded by the regional and national government as well as well as R&D contracts with private companies.

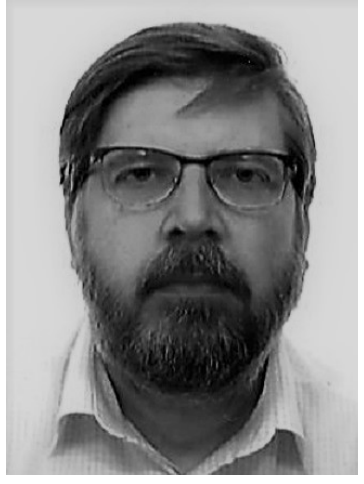

MIGUEL A. VILAR-MONTESINOS received the M.Sc. degree in Electrical Engineering in 1992 from University of Vigo. He started his professional career as an electrical engineer at Astano, subsequently leading to Information Technologies. For the last twelve years he was responsible for Engineering Systems at Navantia and he is currently head of the department of Digitalization Projects. Due to his position, he collaborates with University of A Coruña research on Industry 4.0, Augmented Reality, Internet of Things and RFID systems.